\documentclass[a4paper,oneside,11pt]{article}
\usepackage{amssymb}
\usepackage{graphicx}
\usepackage{multicol} 
\usepackage{floatflt}
\usepackage{color}
\definecolor{rosso}{cmyk}{0,1,1,0.4}
\definecolor{rossos}{cmyk}{0,1,1,0.55}
\definecolor{rossoc}{cmyk}{0,1,1,0.2}
\definecolor{blu}{cmyk}{1,1,0,0.3}
\definecolor{blus}{cmyk}{1,1,0,0.6}
\definecolor{bluc}{cmyk}{1,1,0,0.1}
\definecolor{verde}{cmyk}{0.92,0,0.59,0.25}
\definecolor{verdec}{cmyk}{0.92,0,0.59,0.15}
\definecolor{verdes}{cmyk}{0.92,0,0.59,0.4}

\advance\textheight by \topskip
    \font\tenrsfs=rsfs10 at 11pt
\font\sevenrsfs=rsfs7
\font\fiversfs=rsfs5
\newfam\rsfsfam
\textfont\rsfsfam=\tenrsfs
\scriptfont\rsfsfam=\sevenrsfs
\scriptscriptfont\rsfsfam=\fiversfs
\def\mathscr#1{{\fam\rsfsfam\relax#1}}
\def\Lag{\mathscr{L}}

\large\normalsize
\newcommand{\be}{\begin{equation}}
\newcommand{\ee}{\end{equation}}
\newcommand{\ba}{\begin{array}}
\newcommand{\ea}{\end{array}}

\newcommand{\eV}{{\rm eV}}
\newcommand{\cm}{{\rm cm}}
\newcommand{\km}{{\rm km}}
\newcommand{\meV}{\,{\rm meV}}

\def\Red  {}

\def\Black{}
\def\Blue {}
\newcommand{\eq}[1]{~(\ref{eq:#1})}
\newcommand{\MeV}{\,{\rm MeV}}

\newcommand{\PL}{Phys. Lett.}
\newcommand{\PR}{Phys. Rev.}
\newcommand{\mb}[1]{\mbox{\normalsize\boldmath $#1$}}
\newcommand{\fig}[1]{~\ref{fig:#1}}

\def\circa#1{\,\raise.3ex\hbox{$#1$\kern-.75em\lower1ex\hbox{$\sim$}}\,}
\makeatletter
\def\art{\@ifnextchar[{\eart}{\oart}}
\def\eart[#1]#2#3#4#5#6{{\rm #2}, {#3  #4} {\rm (#6) #5} [#1]}
\def\hepart[#1]#2{{\rm #2, #1}}
\newcommand{\oart}[5]{{\rm #1}, {#2  #3} {\rm (#5) #4}}

\newcounter{alphaequation}[equation]
\def\thealphaequation{\theequation\hbox to
0.6em{\hfil\alph{alphaequation}\hfil}}
\def\eqnsystem#1{
\def\@eqnnum{{\rm (\thealphaequation)}}
\def\@@eqncr{\let\@tempa\relax \ifcase\@eqcnt \def\@tempa{& & &} \or
  \def\@tempa{& &}\or \def\@tempa{&}\fi\@tempa
  \if@eqnsw\@eqnnum\refstepcounter{alphaequation}\fi
\global\@eqnswtrue\global\@eqcnt=0\cr}
\refstepcounter{equation} \let\@currentlabel\theequation \def\@tempb{#1}
\ifx\@tempb\empty\else\label{#1}\fi
\refstepcounter{alphaequation}
\let\@currentlabel\thealphaequation
\global\@eqnswtrue\global\@eqcnt=0 \tabskip\@centering\let\\=\@eqncr
$$\halign to \displaywidth\bgroup \@eqnsel\hskip\@centering
$\displaystyle\tabskip\z@{##}$&\global\@eqcnt\@ne
\hskip2\arraycolsep\hfil${##}$\hfil& \global\@eqcnt\tw@\hskip2\arraycolsep
$\displaystyle\tabskip\z@{##}$\hfil
\tabskip\@centering&\llap{##}\tabskip\z@\cr}
\def\endeqnsystem{\@@eqncr\egroup$$\global\@ignoretrue} \makeatother

\begin{document}
\centerline{hep-ph/0503246 \hfill  \hfill IFUP--TH/2005--06}
\vspace{0.5cm}
\centerline{\LARGE\bf\Red Implications of neutrino data circa 2005}
\medskip\bigskip\Black
  \centerline{\large\bf Alessandro Strumia}\vspace{0.2cm}
  \centerline{\em Dipartimento di Fisica dell'Universit\`a di Pisa and INFN}\vspace{0.4cm}
  \centerline{\large\bf  Francesco Vissani}\vspace{0.2cm}
  \centerline{\em INFN, Laboratori Nazionali del Gran Sasso,
Theory Group, I-67010 Assergi (AQ), Italy}

\bigskip

\Blue\centerline{\large\bf Abstract}

\begin{quote}\large\indent
Adopting the 3 neutrino framework,
we present an updated determination of the oscillation parameters.
We perform a global analysis and develop
simple arguments that give essentially the same result.
We also discuss
determinations of solar neutrino fluxes, capabilities of 
future experiments, tests of CPT, implications for neutrino-less 
double-$\beta$ decay,
$\beta$ decay, cosmology.
\Black
\end{quote}



\begin{table}[h]
$$\begin{array}{|lrlc|}\hline
\hbox{Oscillation parameter}&\multicolumn{2}{c}{\hbox{central value}} &\hbox{$99\%$ CL range}\\  \hline
\color{rossos}
\hbox{solar mass splitting} & \color{rossos}\Delta m^2_{12} ~=
& \color{rossos}(8.0\pm 0.3)\,10^{-5}\eV^2 & \color{rossos}(7.2\div 8.9)\,10^{-5}\eV^2 \\
\color{blus}\hbox{atmospheric mass splitting~}  &\color{blus}|\Delta m^2_{23}| ~=
&\color{blus}~(2.5\pm 0.3) \,10^{-3}\eV^2~ &\color{blus}(1.7\div 3.3) \,10^{-3}\eV^2\\
\color{rossos}
\hbox{solar mixing angle} &\color{rossos} \tan^2 \theta_{12} ~=& \color{rossos}0.45\pm0.05 
&\color{rossos}30^\circ < \theta_{12}<38^\circ \\
\color{blus}\hbox{atmospheric mixing angle} &\color{blus}\sin^2 2\theta_{23} ~=
&\color{blus}  1.02\pm 0.04 &36^\circ <\theta_{23}< 54^\circ\\
\color{verdes}\hbox{`CHOOZ' mixing angle}  &\color{verdes} \sin^2 2\theta_{13}  ~=&
\color{verdes}0\pm0.05 &\color{verdes} \theta_{13}<10^\circ\\ \hline
\end{array}$$\vspace{-4mm}
$$\begin{array}{|ccc|cc|}\hline
\hbox{non-oscillation} &\hbox{probed} &\hbox{experimental} &  \hbox{$99\%$ CL range} &  \hbox{$99\%$ CL range}\\
\hbox{parameter}&\hbox{by}&\hbox{limit  at $99\%$ CL}& \hbox{normal hierarchy} &   \hbox{inverted hierarchy}\\ \hline
\hbox{$ee$-entry of $m$}&\hbox{$0\nu2\beta$} &m_{ee}< 0.38\,h\,\eV&(1.1\div 4.5)\meV  & (12\div 57)\meV\\
\hbox{$(m^\dagger m)^{1/2}_{ee}$}&\hbox{$\beta$-decay} & m_{\nu_e}<2.0\,\eV& (4.6\div10)\meV &(42\div 57)\meV \\
\hbox{$m_1+m_2+m_3$} &\hbox{cosmology}& m_{\rm cosmo}<0.94\,\eV &(51\div 66)\meV& (83\div 114)\meV\\ \hline
\end{array}$$
\caption{\em Summary of present
information on neutrino masses
and mixings from oscillation data (upper rows)  and inferences and limits
on non-oscillation probes (lower rows)
A $99\%$ C.L.\ range is a $2.58\sigma$ range.
Some $0\nu2\beta$ data are controversial, and  $h\sim 1$ parameterizes uncertain nuclear matrix elements.
\label{tab1}}
\end{table}

\begin{figure}[h]\vspace{-9mm}
$$\hspace{-8mm}\includegraphics[width=168mm]{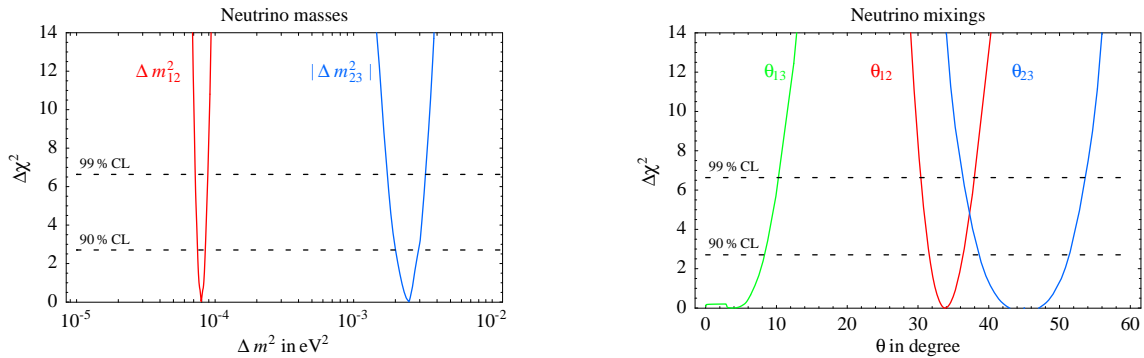} $$
\vspace{-11mm}
  \caption[]{\em Summary of present
information on neutrino masses
and mixings from oscillations.
\label{fig:panta}}
\end{figure}

\newpage

\noindent
The most plausible extension of the Standard Model that allows to interpret a wealth of neutrino data~\cite{Chlorinelast,Galliumlast,Gallex,SAGE, SKlast, SNOlast, SNOsaltfinal, KL2004,SKI,Macro,K2K}
consists in adding a Majorana mass term for neutrinos, 
\begin{equation} \Lag = \Lag_{\rm SM} + \frac{1}{2}  ({\nu} \cdot m \cdot \nu + \hbox{h.c.})\,\qquad
m=V^*\cdot  \mbox{diag}( m_1,\ m_2\ e^{2i\alpha},\ m_3\ e^{2i\beta}) \cdot V^\dagger.
\end{equation}
For normal mass hierarchy we order the neutrino masses $m_{1,2,3}\ge 0$ as
$m_1<m_2<m_3$,  whereas for inverted mass hierarchy we choose $m_3<m_1<m_2$. 
The neutrino mixing matrix is
\begin{equation}
V=R_{23}(\theta_{23})\
\mbox{diag}(1,e^{i\phi},1)\ R_{13}(\theta_{13}) 
\ R_{12}(\theta_{12}).\end{equation}
A similar result holds in the case of Dirac masses,
with the difference that the number of physical parameters decreases from 9 to 7:
the Majorana phases $\alpha$ and $\beta$   can be reabsorbed by field redefinitions.

Data on neutrino oscillations fix $\theta_{12},\theta_{23},\Delta m^2_{12}$
and $|\Delta m^2_{23}|$ where $\Delta m^2_{ij}\equiv m_j^2-m_i^2$.
As discussed later (sections \ref{sec:sol} and \ref{sec:atm}) 
our present knowledge of oscillation parameters 
is approximatively summarized in the upper rows of table~\ref{tab1}.
The uncertainties are almost Gaussian in the chosen variables;
fig.\fig{panta} shows the  full $\chi^2$ functions.
Correlations among parameters are ignored because negligible,
with the exception that the upper 
bound on $\theta_{13}$ depends on $|\Delta m^2_{23}|$.
While $\theta_{12}$ is rather precisely measured, 
the other 2 mixing angles have large uncertainties. 
Planned long-baseline oscillation experiments can strongly 
improve on $\Delta m^2_{23}$
and (if $\theta_{13}\circa{>}1^\circ$) measure  $\theta_{13}$, 
the phase $\phi$, and determine which type of mass hierarchy
is realized in nature.

Oscillation experiments, however, are insensitive to 
the absolute neutrino mass scale (say, the mass of the 
lightest neutrino) and to the 2 Majorana phases $\alpha$ and $\beta$.
Other types of experiments can study some of
these quantities and the nature of neutrino masses. They are:
$\beta$-decay experiments, that in good approximation probe
$m_{\nu_e}^2\equiv (m\cdot m^\dagger)_{ee}=\sum_i |V_{ei}^2| m_i^2$; 
neutrino-less double-beta decay ($0\nu2\beta$) experiments, that probe 
the absolute value of the Majorana mass
$m_{ee} \equiv \sum_i V_{ei}^2 m_i$; 
cosmological observations (Large Scale Structures and anisotropies in the Cosmic Microwave Background), 
that in good approximation probe
$ m_{\rm cosmo}\equiv \Omega_\nu h^2\cdot 93.5\eV= \sum_i m_i$. 
Only neutrinoless double beta decay
($0\nu 2\beta$) probes the Majorana nature of the mass.
The values $|m_{ee}|, m_{\nu_e}, m_{\rm cosmo}$
 are unknown, but can be partially inferred from oscillation data.
Table~\ref{tab1} shows our results, discussed in section~\ref{sec:bb},
in the limit where the mass of the lightest neutrino is negligible.
In the opposite limit neutrinos are quasi-degenerate and $|m_{ee}|, m_{\nu_e}, m_{\rm cosmo}$ can be arbitrarily large.

{}From the point of view of 3 massive neutrinos, 
it is natural to divide in three parts 
a discussion of 
the present situation and of the 
perspectives of improvements, namely: 
\begin{itemize}
\item[1.] Oscillations with `solar' frequency, 
that tell $\Delta m^2_{12}$ and $\theta_{12}$ and give a sub-dominant constraint on $\theta_{13}$.
In section~\ref{sec:sol} we discuss solar and reactor neutrino
experiments, showing that the program of measurement of parameters 
is well under way (if not accomplished), and discussing other interesting related goals.

\item[2.] Oscillations with `atmospheric' frequency, that
tell $\Delta m^2_{23}$, $\theta_{23}$ and $\theta_{13}$, are discussed in
section~\ref{sec:atm}.

\item[3.] Non-oscillation experiments, that can tell the absolute neutrino mass.
In section~\ref{sec:bb} we discuss the present status 
and assess the implications of the existing 
information on neutrino oscillations for these experiments,
particularly for $0\nu2\beta$.
We conclude by commenting on the recent claim of evidence for 
this transition~\cite{Klapdor, KlapdorLast}.

\end{itemize}
We assume the 3 neutrino framework because it is
plausible, well defined, restrictive and compatible with data. 
However it is just an assumption, and
before proceeding we recall some alternatives.
The most plausible one is the presence of extra light fermions
(`sterile neutrinos') or bosons, which can manifest in many ways.
Going to rather exotic scenarios, Lorentz  or CPT invariance (here studied in fig.\fig{CPT2005}) might be violated in neutrinos,
that might have anomalous interactions (gauge couplings, or magnetic moments, or else), 
might not obey the Pauli principle, etc, etc, etc.
Present solar and atmospheric data cannot be explained by these alternatives,
which however might be present as sub-leading effects on top of
oscillations among active neutrinos, such that
our determinations of active oscillation parameters
would need model-dependent modifications.
The present data  do not give 
any clear indication for extra effects, but contain some anomalous hints.
Most notably, the LSND anomaly~\cite{LSND} is not compatible with the 3 neutrino context we assume.

\section{Oscillations with solar frequency\label{sec:sol}}
A few years ago the solar anomaly rested on
global fits that combined solar model predictions with
a few measurements of solar neutrino rates.
In recent times, the situation changed.
As prospected  in \cite{which}
(written a few years ago, while 
sub-MeV solar experiments were discussed as a tool for discriminating LMA from 
LOW, SMA, QVO,\ldots),
SNO, KamLAND,  {\sc Borexino}
had in any case the capability to identify the true solution of the solar anomaly
and make precision measurements of the oscillation parameters.
This is where we are now. From the point of 
view of the determination of the oscillation  parameters, 
solar neutrino experiments are in a more advanced 
stage than atmospheric experiments, as clear from fig.~\ref{fig:panta}.
KamLAND and SNO play the key r\^ole in the determination 
for $\Delta m^2_{12}$ and $\theta_{12}$ respectively,
and almost achieved the $2.5\%$ and $10\%$ accuracy
prospected in~\cite{which}.
In section~\ref{sec:fit} we present a global solar fit,
and in section~\ref{sec:gl} we show that it is dominated by very simple and robust inputs.
In section~\ref{sec:PeeLow} we extract from data the survival probability of
low-energy neutrinos, and use it to study how solar data mildly restrict $\theta_{13}$.
In section~\ref{sec:bx} we reassess the interest in proceeding 
with low-energy solar neutrino experiments.
In section~\ref{sec:Phi} we discuss how well present solar data
determine solar neutrino fluxes and discuss the impact of {\sc Borexino}.
In all cases we extract  from simple arguments the main general results,
that we compare with `exact' results of global fits performed in specific cases.

\subsection{Updated fit of solar and reactor neutrino data\label{sec:fit}}
We begin by presenting a 
global fit of solar and reactor neutrino data assuming oscillations
among active neutrinos with negligible $\theta_{13}$. We include
 \begin{itemize}

\item The final SNO CC, NC, ES rates measured during day and during night
without~\cite{SNOlast} and with~\cite{SNOsaltfinal}  salt, that gives enhanced NC sensitivity.

\item The Super-Kamiokande ES spectra~\cite{SKlast}.

\item The Gallium rate, $R_{\rm Ga} = (68.1\pm3.7)\,{\rm SNU}$, obtained averaging
the most recent  SAGE data with the final {\sc Gallex} and GNO data~\cite{Galliumlast,Gallex,SAGE}.

\item The Chlorine rate~\cite{Chlorinelast},
$R_{\rm Cl} = (2.56 \pm 0.23)\,{\rm SNU}$.

\item The KamLAND reactor anti-neutrino
data with prompt energy higher than $2.6\MeV$~\cite{KL2004}.

\end{itemize}
Solar model predictions and uncertainties~\cite{BP} are revised
including the recent measurement of the
$^{14}$N$(p,\gamma)^{15}$O nuclear cross section~\cite{LUNA},
which reduces the predicted CNO fluxes by roughly $50\%$.

\begin{figure}
$$
\includegraphics[width=8cm]{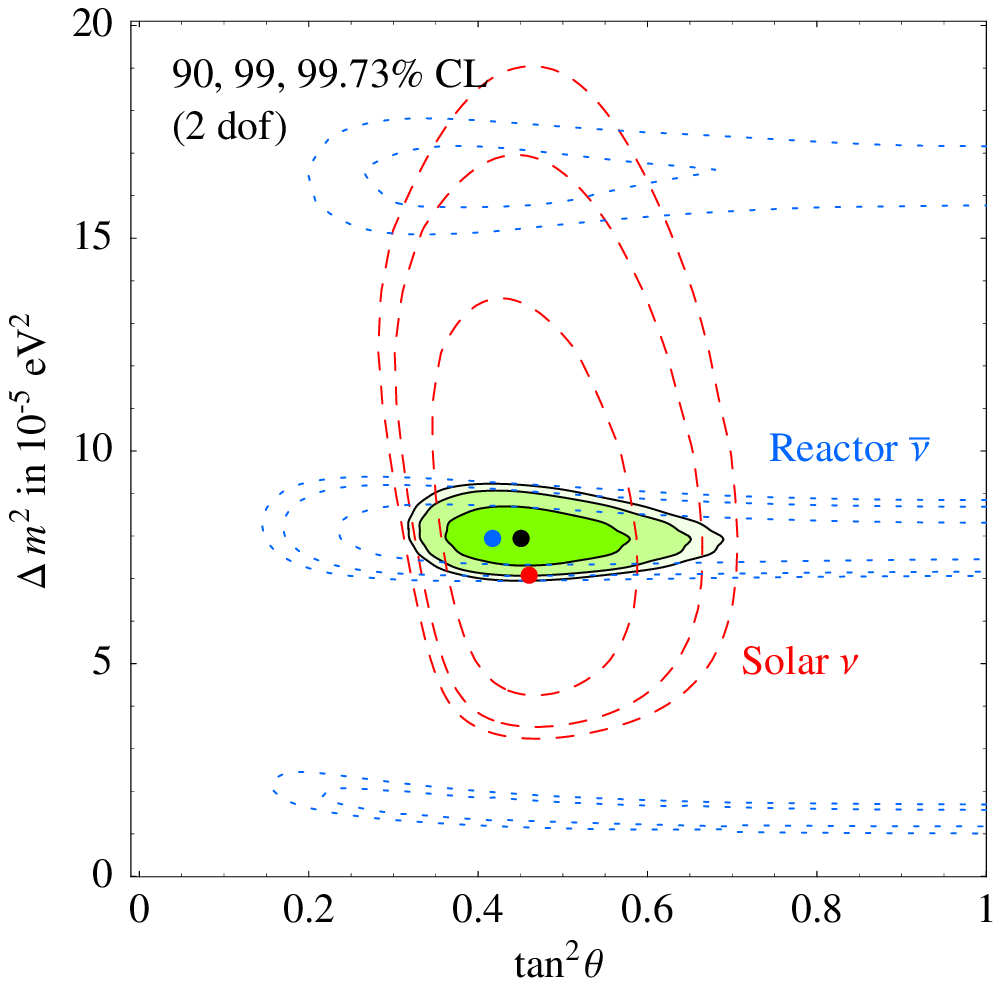}\qquad 
\includegraphics[width=8cm]{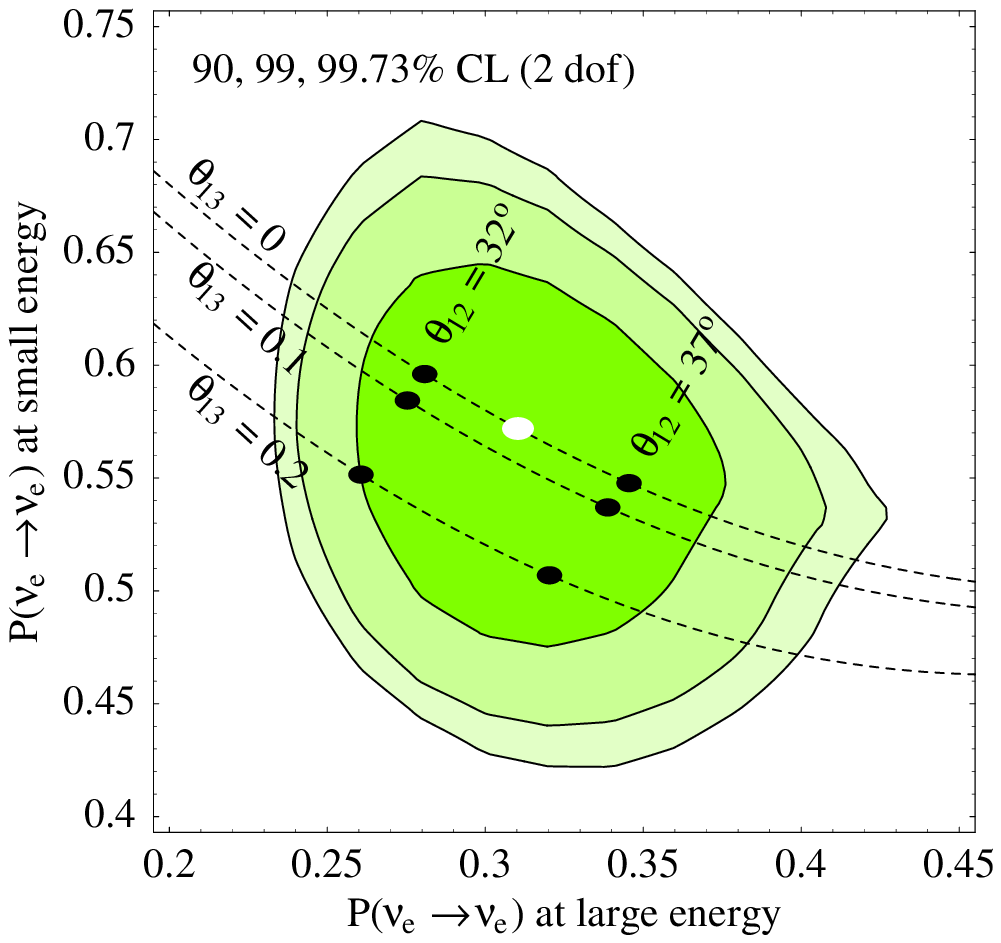}$$
\vspace{-5mm}
\caption[x]{\label{fig:2005}\em 
Best-fit regions at $90,~99$ and $99.73\%$ {\rm CL}.
Fig.~\ref{fig:2005}a assumes CPT invariance and combines
solar $\nu$ data (dashed red contours) with reactor $\bar{\nu}$ data (dotted blue contours).
In fig.~\ref{fig:2005}a we show how data determine the high- and low-energy limits
of $P(\nu_e\to\nu_e)$, as precisely described in the text,
}
\end{figure}

The result of our oscillation fit is shown in fig.~\ref{fig:2005}a, where
the best fit point is marked with a dot.
The $1\sigma$ (i.e.\ $68\%$ C.L.) and the $99\%$ C.L. (i.e.\ $2.58\sigma$) ranges for the single parameters (1 dof) are
summarized in table~\ref{tab1}.
The total evidence for an effect
is now about 12$\sigma$ in solar $\nu$ data and about 6$\sigma$ in KamLAND $\bar\nu$ data.
Fig.~\ref{fig:2005}a also shows separately the fit of solar data (dashed red contours)
and the fit of KamLAND data (dotted blue contours).

We note in passing that, according
to the Pearson $\chi^2$ goodness-of-fit (gof) employed by many old
global analyses, today the LOW solution (with $\Delta m^2_{12}\approx 10^{-7}\eV^2$)
still has a good gof.
This does not mean that LOW  is compatible with data, but 
happens because the Pearson test is not a statistically powerful
gof test when $\hbox{dof}\gg 1$~\cite{sunfitsalt}: global fits of solar data input
roughly $100$ experimental inputs.
As we now discuss, the output depend almost only on a few pieces of data.

\begin{figure}[t]
$$\includegraphics[width=8cm]{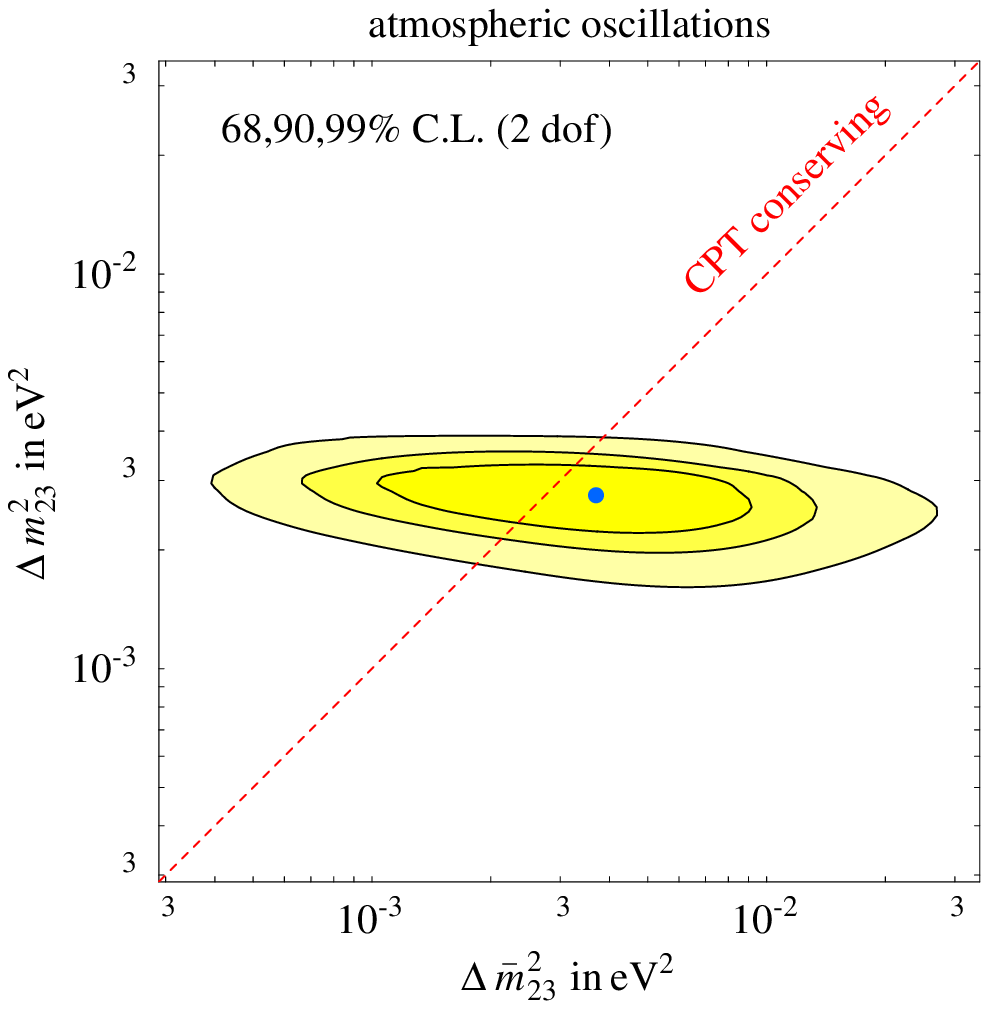}\qquad
\includegraphics[width=8cm]{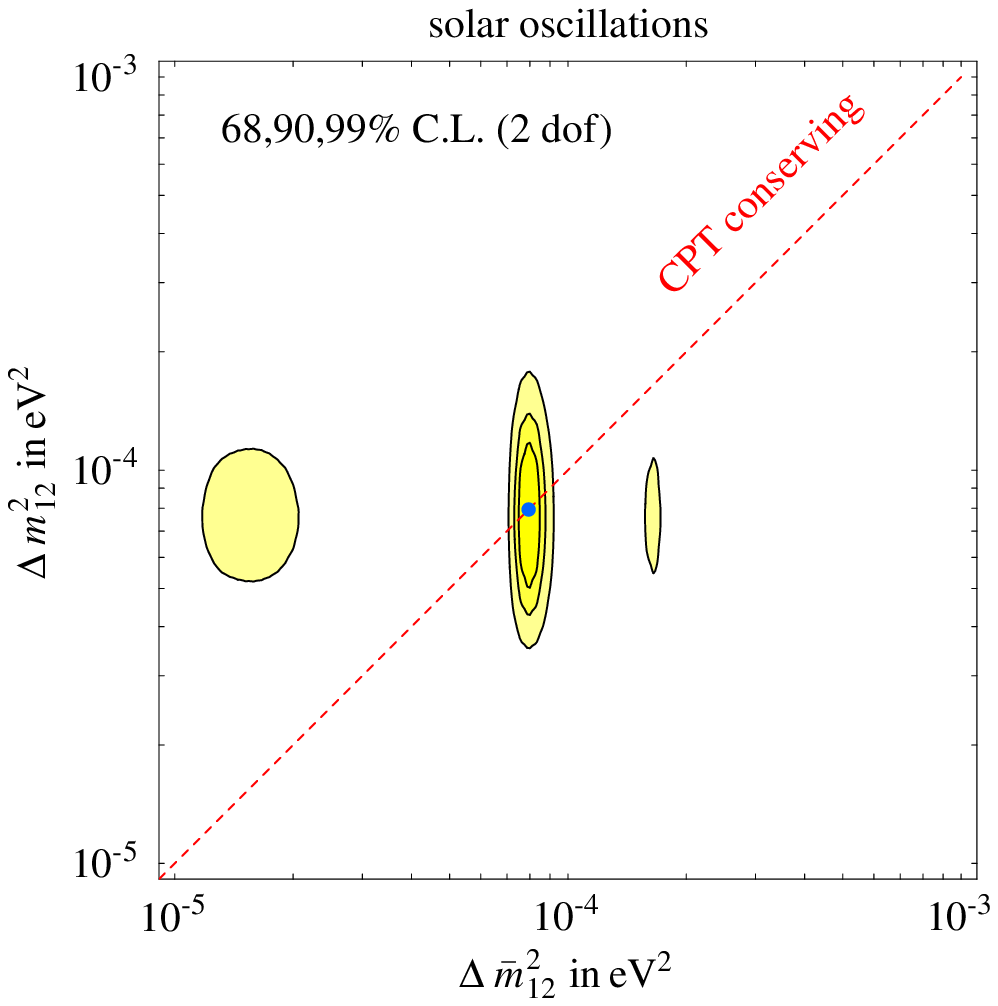}$$
\caption[x]{\label{fig:CPT2005}\em 
Test of CPT-violating neutrino masses.
We show  the separate fit for $\Delta m^2$ in neutrinos and $\Delta\bar{m}^2$ in
anti-neutrinos, marginalized with respect to the mixing angles
$\theta_{12}$ and $\theta_{23}$.
The atmospheric fit includes data from SK, K2K, Macro.
The solar fit includes data from SNO, SK, {\sc Gallex}, {\sc Sage}, {\sc Homestake}, KamLAND.
These plots update the original results in fig.s 5 and 6 of~\cite{CPT}}\end{figure}

\subsection{The meaning of `global fits'\label{sec:gl}}
Our results are based on a careful global fit.
We now point out how a simple approximate
analysis is sufficient to get results practically 
equivalent to those of global fits.

{\em The solar mass splitting} $\Delta m^2_{12}$ is directly determined by the position of the oscillation dips at
KamLAND, with negligible
contribution from solar experiments.
(More precisely, this will be rigorously true in the future.
For the moment solar data are needed to eliminate spurious solutions
mildly disfavored by KamLAND data, as illustrated in fig.~\ref{fig:2005}a.
Solar data have practically no impact around the global minimum of
the $\chi^2$, and consequently on the determination of $\Delta m^2_{12}$ as reported in table~\ref{tab1}).

\bigskip

{\em The solar mixing angle} is directly determined by SNO measurements of NC and CC solar Boron rates.
Assuming  flavour conversions among active neutrinos, SNO implies\footnote{SNO measured
$ \Phi(\nu_{e,\mu,\tau}) = (4.9\pm0.3)~10^6/\cm^2\sec$ and
$\Phi(\nu_{e}) =(1.74\pm0.08)~10^6/\cm^2\sec$.
Each measurement was first performed using
heavy water (CC and NC events mainly  distinguished by their energy spectrum)
and later with salt heavy water
(distinction relies on event isotropy).
The measurements performed with these
different experimental techniques agree.
Taking into account the SK measurement of the ES rate,
$\Phi(\nu_{e})  + 0.155 \Phi(\nu_{\mu,\tau}) = (2.35\pm0.06)~10^6/\cm^2\sec $,
the SNO measurement of the ES rate,
$\Phi(\nu_{e})  + 0.155 \Phi(\nu_{\mu,\tau}) = (2.36\pm0.19)~10^6/\cm^2\sec $
and the solar model prediction
$\Phi(\nu_{e,\mu,\tau}) = (5.05\pm 0.86)~10^6/\cm^2\sec $
would only marginally improve the measurement of $\langle P(\nu_e\to\nu_e)\rangle$ to
$\langle P(\nu_e\to\nu_e)\rangle=0.360\pm0.028$.}
$$\langle P(\nu_e\to\nu_e) \rangle \equiv  {\Phi(\nu_{e}) }/{\Phi(\nu_{e,\mu,\tau}) }=0.357\pm0.030.$$
This should be compared with the theoretical prediction for $\langle P(\nu_e\to\nu_e)\rangle$,
given by a simple expressions that does not depend on the solar density profile
because LMA oscillations are almost completely 
adiabatic.
At $E_\nu\gg \MeV$ matter effects
dominate such that $\nu_e$ produced around the center of the sun 
coincide with the $\nu_2$ eigenstate in matter and exit as the
$\nu_2$ eigenstate in vacuum, so that $P(\nu_e\to\nu_e)\simeq \sin^2\theta_{12}$.
In the energy range explored by SNO, matter effects at the production region
are not fully dominant,
such that the above approximation gets slightly corrected to\footnote{The 
factor giving the correction to $P(\nu_e\to\nu_e) = \sin^2\theta_{12}$
ranges between $1.1$ and $1.2$ within the 
present best-fit region at $90\%$ CL.}
\begin{equation}\label{eq:1.15}
\langle P(\nu_e\to\nu_e) \rangle \approx 1.15\sin^2\theta_{12}
\qquad\hbox{so that}\qquad
 \tan^2\theta_{12}=0.45\pm0.05
 \end{equation}
 which agrees with the results of the global analysis in table~\ref{tab1},
 both in the central value and in its uncertainty.

Notice that the only solar model input that enters
our approximate determination of solar oscillation parameters
is the solar density around the center of the sun,
that controls the $15\%$ correction to $\langle P(\nu_e\to\nu_e)\rangle$ in eq.\eq{1.15}.
This correction factor is comparable 
to the $1\sigma$ uncertainty in $\langle P(\nu_e\to\nu_e)\rangle$:
indeed the associated increase of $P(\nu_e\to\nu_e)$ at smaller $E_\nu$ has not been observed
in SNO and SK spectra.
For the reasons explained above the `solar model independent fit' of~\cite{BS}
gives now a result almost identical to the standard fit,  so that we do not show the update of this result.

Global fits 
remain still useful for testing if the pieces of data not included
in our simplified analysis (that have a minor impact in the standard fit)
contain statistically significant indications for
new physics beyond LMA oscillations.
At the moment the answer is no.
E.g., fig.~\ref{fig:CPT2005}b  updates the CPT-violating solar fit of~\cite{CPT}:
the best-fit region includes the CPT-conserving limit (diagonal dotted line).

\subsection{Effects of $\theta_{13}$ and low-energy neutrinos}\label{sec:PeeLow}
\begin{figure}
$$\includegraphics[width=7.5cm]{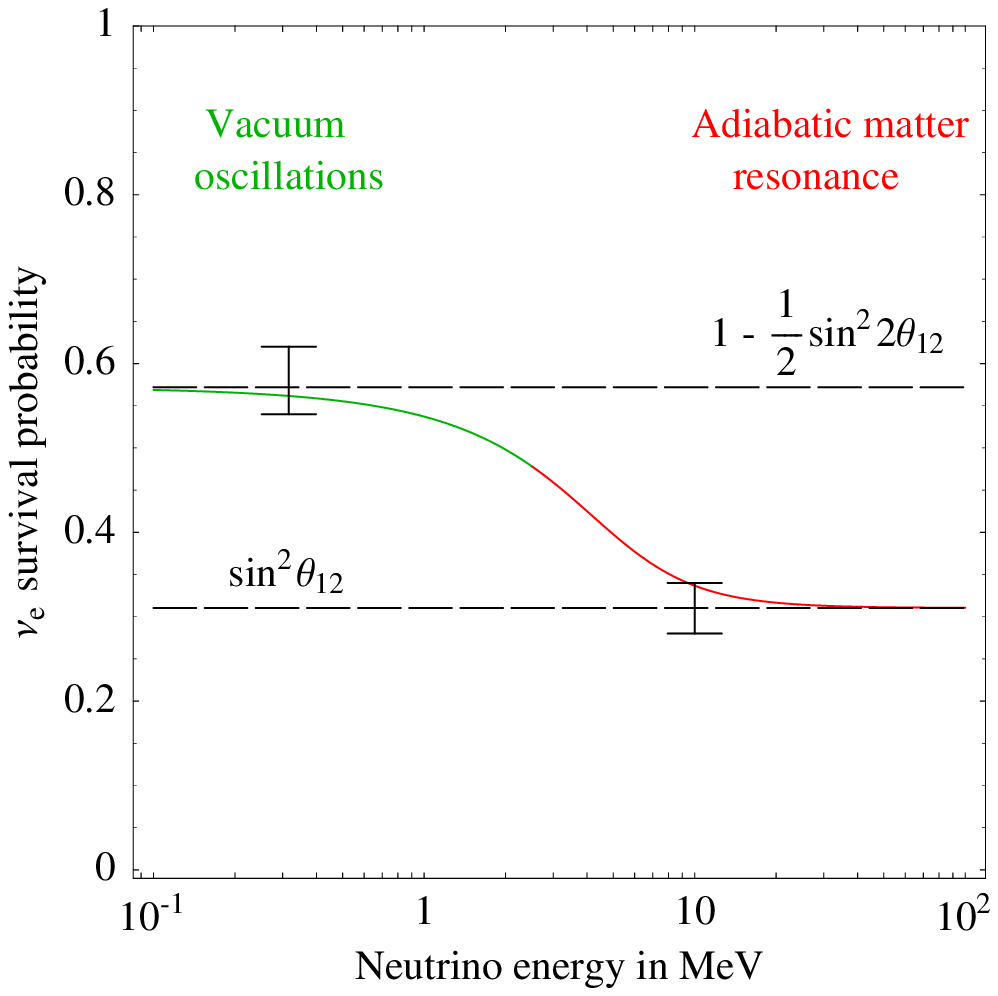}\qquad\qquad
\includegraphics[width=7.5cm]{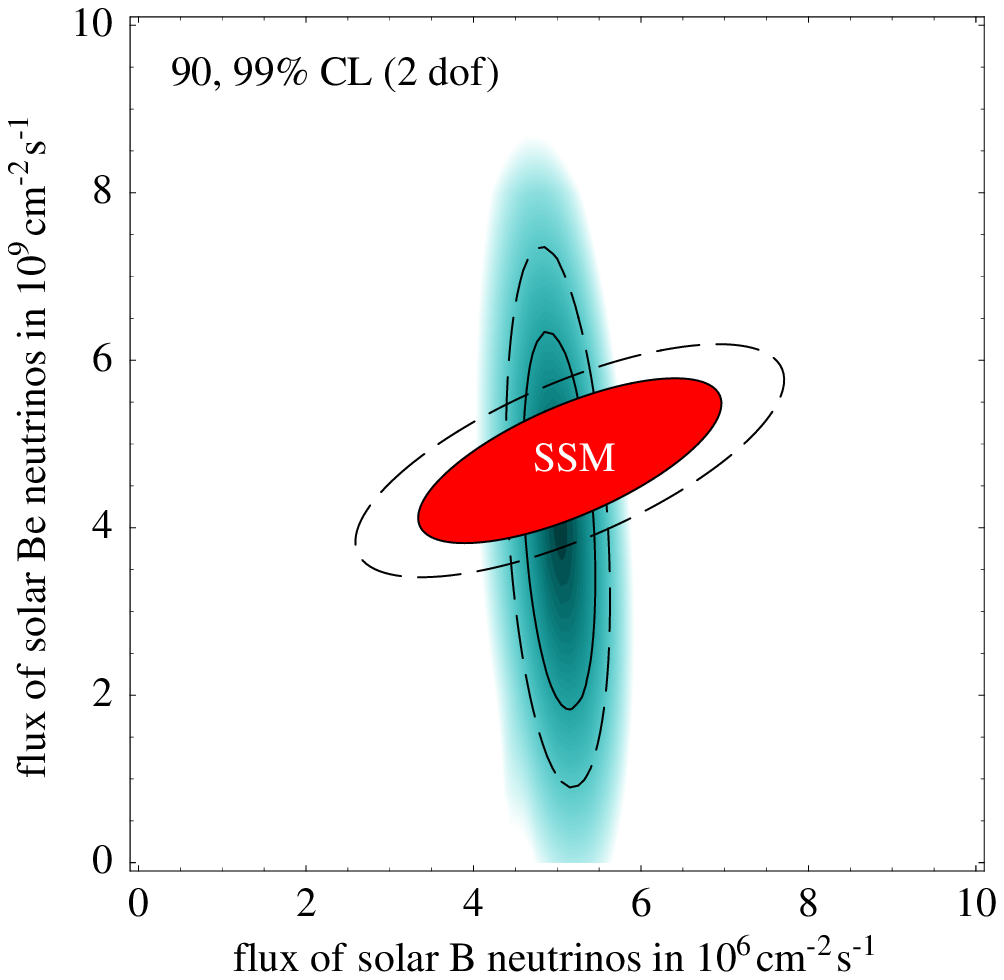}$$
\parbox{0.47\textwidth}{\caption[x]{\em \label{fig:OscLMA}
The energy-dependent survival probability
 predicted by LMA,
and how experimental data restrict the low-energy and
high-energy limits of $P(\nu_e\to \nu_e,E_\nu)$.}}\hspace{0.06\textwidth}
\parbox{0.47\textwidth}{\caption{\label{fig:BeB2005}\em 
Best-fit regions at $90,~99\%$ {\rm CL} (2 dof) 
 for Boron and Beryllium fluxes
 from a global solar-model-independent fit of solar data.
The horizontal ellipse shows solar model predictions. 
}}
\end{figure}

At low $E_\nu$ matter effects are negligible and
the survival probability is given by averaged vacuum oscillations.
LMA oscillations with $\theta_{13}=0$ predict the low-energy limit of
$P(\nu_e\to \nu_e)$ in terms of its high-energy limit as\footnote{The transition between the two regimes
proceeds at $E_\nu\sim \hbox{few}\,\MeV$, and the high-energy regime is approached
for $E_\nu \circa{>} 20\MeV$.  
At higher energies solar matter effects become comparable to the
atmospheric mass splitting, giving corrections to solar neutrino rates
proportional to $\theta_{13}^2$.}
\begin{equation}\label{eq:pred}
P(\nu_e\to\nu_e,\hbox{small $E_\nu$}) = 1- 2P(\nu_e\to\nu_e,\hbox{large $E_\nu$}) +
2P(\nu_e\to\nu_e,\hbox{large $E_\nu$})^2.
\end{equation}
A non zero $\theta_{13}$ as well as new physics beyond neutrino masses
allow to avoid this prediction.
It is therefore interesting to extract these two ideal observables from data.
As discussed in the previous section
$P(\nu_e\to\nu_e,\hbox{large $E_\nu$})$ is presently dominantly determined by SNO.
The low energy limit of $P(\nu_e\to\nu_e)$ is presently dominantly determined by
Gallium data and can be extracted by a simple approximate argument~\cite{which,SAGE}.
  Subtracting from the total Gallium rate 
\begin{equation}\label{eq:Ga}
(68.1\pm 3.7) \,\hbox{SNU}=R_{\rm Ga} =
R_{pp,pep}^{\rm Ga} +  R_{\rm CNO}^{\rm Ga} +  R_{^7{\rm Be}}^{\rm Ga} + R_{^8{\rm B}}^{\rm Ga} 
\end{equation}
 its  $^8$B contribution 
 (as directly measured by SNO via CC, $R_{^8{\rm B}}^{\rm Ga} = 4.3\pm 1\,\hbox{SNU}$)
 and regarding all remaining fluxes as low energy ones, suppressed by
$P(\nu_e\to\nu_e,\hbox{small $E_\nu$})$, determines it to be $0.57\pm 0.03$.
Alternatively,
by subtracting also the intermediate-energy CNO and Beryllium fluxes, one gets
$P(\nu_e\to\nu_e,\hbox{small $E_\nu$})=0.58\pm0.05$.
We here included only the error on the Gallium rate, which is the dominant error.
This rough analysis shows that the result only mildly depends on how one deals with intermediate energy neutrinos, and on
model-dependent  details of the intermediate region,
thereby suggesting the following general result:
\begin{equation}\label{eq:PeeVO}
P(\nu_e\to\nu_e,\hbox{large $E_\nu$}) =
0.31\pm0.03,\qquad
P(\nu_e\to\nu_e,\hbox{small $E_\nu$}) = 0.58\pm0.04.
\end{equation}
As illustrated in fig.\fig{OscLMA} at $E_\nu =1\MeV$ ($10\MeV$) the difference
between the exact LMA profile of $P(\nu_e\to\nu_e)$
and its low energy (high energy) limit is smaller than the
$1\sigma$ uncertainties in eq.\eq{PeeVO}.
High energy neutrinos have been dominantly measured by SNO at energies around $10\MeV$.

In order to establish the validity of eq.\eq{PeeVO} we compare it
with the `exact' result of a global analysis of solar data: to perform it we must 
abandon generality and focus on a specific mechanism
that allows to avoid the LMA prediction of eq.\eq{pred}.
We consider the case of a non vanishing $\theta_{13}$ (see also~\cite{theta13sun}),
which gives
\begin{eqnsystem}{sys:PeeLowHi}\label{eq:PeeHi}
P(\nu_e\to\nu_e,\hbox{large $E_\nu$}) &=&
 \sin^4\theta_{13}+\cos^4\theta_{13}\sin^2\theta_{12},\\
P(\nu_e\to\nu_e,\hbox{small $E_\nu$}) &=& \sum_i |V_{ei}|^4=
 \sin^4\theta_{13}+\cos^4\theta_{13}\bigg[1-\frac{1}{2}\sin^2 2\theta_{12}\bigg].
 \label{eq:PeeLow}
\end{eqnsystem}
In this way the equality in eq.\eq{pred} gets replaced by a $\le$ inequality.
In order to access also the other region we analytically continue
to imaginary $\theta_{13}$, where $\cos^2\theta_{13}>1$.
(This is analogous but less usual than allowing $\sin^2 2\theta_{23}>1$ in the atmospheric fit).
The global fit is performed by keeping fixed $\Delta m^2_{12}$ at the central value suggested by KamLAND,
because a variation of $\Delta m^2_{12}$
within the range in table~\ref{tab1} negligibly affects solar data.
This is the key extra input provided by KamLAND; a more complicated 
solar plus KamLAND global fit would give the same result.

The result of the global solar fit performed in this specific context is shown
in  fig.\fig{2005}b, and agrees with the semi-quantitative general result  of eq.\eq{PeeVO}.
We see that the LMA prediction in eq.\eq{pred} is well compatible with data.
The constraint on $\theta_{13}$ provided by solar data (subdominant with respect to the constraint from CHOOZ and atmospheric data)
is included in the global analysis summarized in table~\ref{tab1} and fig.\fig{panta}.

We performed more global fits considering a few other ways of avoiding the LMA prediction of
eq.\eq{pred}: in each case the allowed region looks like a potato
similar to the one in fig.\fig{2005}b.
Different fits give allowed regions  with
sizes and shapes that vary roughly as much as potatoes vary.
For example we tried to linearly distort
 the $P(\nu_e\to \nu_e,E_\nu)$ profile predicted by LMA for $\theta_{13}=0$ as 
$P(\nu_e\to \nu_e,E_\nu)=
\sin^2\theta_{12}+\lambda[P_{\rm LMA}(\nu_e\to \nu_e,E_\nu)-\sin^2\theta_{12}]$.
This is somewhat different than considering a $\theta_{13}\neq 0$, because
matter effects depend on $\theta_{13}$,  that therefore does not act as a linear distortion.


The main point is that the approximate general result of eq.\eq{PeeVO} fairly summarizes
the variety of exact results obtained by performing global fits in presence of 
different  mechanism that distort the prediction of eq.\eq{pred}
without introducing new notable features at intermediate energies.
Therefore eq.\eq{pred} is a useful semi-quantitative way
of summarizing our present knowledge of $P(\nu_e\to \nu_e)$.
Its behavior at  intermediate energies $1\MeV\circa{<}E_\nu\circa{<}10\MeV$
is basically unknown.

\subsection{Low energy neutrinos and B{\small}{OREXINO} \label{sec:bx}}
Can future sub-MeV solar neutrino experiments improve on oscillation parameters?
This question was answered in~\cite{which},
and we do not have much to add; for recent 
works on the subject see~\cite{roadmaps}. 
We recall here the main points. Low energy solar neutrinos 
do not experience the MSW resonance in the sun:
their survival probability is therefore 
given by the averaged vacuum oscillations
expression of eq.\eq{PeeLow}.
This means that, in first approximation, sub-MeV experiments have nothing 
to tell about $\Delta m^2_{12}$, but could give information on $\theta_{12}$
by measuring $P(\nu_e\to\nu_e,\hbox{small $E_\nu$})$.
Within the 3 neutrino context this same survival probability of eq.\eq{PeeLow} is also directly measured by
reactor $\bar\nu_e$ experiments, such as KamLAND.
KamLAND can achieve a determination of $P(\nu_e\to\nu_e,\hbox{small $E_\nu$})$
competitive with the solar result in eq.\eq{PeeVO}.
A future reactor experiment with baseline $\sim 50\km$ appropriate for observing the first
oscillation dip could achieve an error $(2\div 3)$ times lower than
the present error in eq.\eq{PeeVO}.
Such reduced uncertainty is already achieved today, if one trusts the LMA prediction of eq.\eq{pred}
and employs it to infer the low-energy limit of $P(\nu_e\to \nu_e)$ from SNO data:
\begin{equation}\label{eq:PeeLMA}
P(\nu_e\to\nu_e,\hbox{small $E_\nu$}) = 1-\frac{1}{2}\sin^22\theta_{12}=
0.57\pm 0.02\qquad\hbox{(LMA prediction)}.
\end{equation}
For simplicity we here assumed $\theta_{13}=0$, because its value is not a relevant issue:
 if $\theta_{13}$ is large enough to have a sizable effect, long-baseline experiments will see and measure it
 so well that only the central value of eq.\eq{PeeLMA} (but not its  uncertainty) has to be changed.

The precision of sub-MeV experiments is 
ultimately limited by the $1\%$ solar model uncertainty on the $pp$ flux.
Although it presently seems unrealistic to aim at measuring the $pp$ rate with this level of accuracy, 
the above discussion shows that such ultimate precision should be achieved, 
if sub-MeV experiments want to contribute significantly to the determination of $\theta_{12}$.
Going beyond the $3\nu$ framework, sub-MeV experiments can make important
searches for new physics beyond neutrino masses.
The reason is that high energy solar neutrinos, detected by SNO,
are almost pure $\nu_2$ (the neutrino mass eigenstate with mass $m_2$),
so that low energy experiments are needed to probe new physics that dominantly affects $\nu_1$.
This argument applies e.g.\ in presence of an extra sterile neutrino in wide ranges
of its oscillation parameters~\cite{sterile}.

\bigskip

If LMA is the end of the story the near-future experiment {\sc Borexino} 
will not improve the determination of oscillation parameters
and should observe no anomalous day/night nor seasonal variation.
(KamLAND can also be converted into a solar neutrino experiment).
We comment on the impact of 
 {\sc Borexino} (and eventually KamLAND)
from the point of view of new physics.
As discussed below, in presence of generic 
new physics, existing data poorly constrain the Beryllium rate,
so that large deviations from the LMA prediction are possible.
The rate measured by {\sc Borexino} can be modified by new 
physics that mostly affects intermediate energy neutrino.
Rather than performing a detailed analysis of one 
specific source of new physics,
out purpose will be to study this possibility in an 
approximate but sufficiently general way.
To this end, we notice that in {\sc Borexino} 
the effects of unspecified new physics 
dominantly manifest as an anomalous flux of Beryllium neutrinos
({\sc Borexino} also has  a minor sensitivity 
to $pep$ neutrinos~\cite{Borexinopep}).
Therefore, in the next section we study 
how well present data restrict the Beryllium rate.

%

%

\subsection{Solar neutrino fluxes}\label{sec:Phi}
An eventual deviation of solar neutrino fluxes from solar model LMA predictions
could be due to new physics in neutrinos beyond LMA, 
or to (new) physics not included in solar models.
For concreteness we focus on the second case, and study how well 
present solar neutrino data determine solar neutrino fluxes,
updating the results of~\cite{BS}. Apparently a detailed global 
fit seems needed to address this issue, but --- once again --- 
one can answer to this question in a simple way.
Following~\cite{BS} the main points  to consider are the following ones:
\begin{itemize}
\item[1.] The luminosity constraint allows to precisely predict the $pp$ fluxes.
Since no experiment so far is sensitive to deviations compatible with the luminosity constraint,
we can essentially set $pp$ fluxes to their solar model value.

\item[2.] SNO measured the Boron flux.

\item[3.] Only two kind of experiments,
Gallium and Chlorine, have measured low-energy neutrino fluxes.
Therefore, data only constrain two linear combinations of low-energy fluxes.

\item[4.]
The Chlorine experiment has a poor sensitivity to low energy neutrinos:
after subtracting the $\sim80\%$ Boron contribution to the Chlorine rate, 
as directly measured by SNO via CC,
the residual low-energy contributions to the
Chlorine rate is just about $2\sigma$ above zero.

\end{itemize}
Therefore the Chlorine rate carries so little information on low energy fluxes,
that our present knowledge on low-energy fluxes is well summarized by a single number: 
their contribution to the Gallium rate.
Restarting from eq.\eq{Ga}, we now subtract from the total Gallium rate $R_{\rm Ga}$ its  $^8$B contribution 
and its $pp,pep$ contributions
(as predicted by solar models and LMA oscillations, see eq.\eq{PeeVO}: $R_{pp,pep}^{\rm Ga} =41.3\pm1.5$), obtaining
 \begin{equation}\label{eq:CNOBe}
 (22.5 \pm 4)\, \hbox{SNU} =R_{\rm CNO}^{\rm Ga} +  R_{^7{\rm Be}}^{\rm Ga}  =
\frac{4.0 \Phi_{^7{\rm Be}} + 4.6 \Phi_{\rm CNO}}{10^{9}/\cm^2{\rm s}} \,\hbox{SNU}.
 \end{equation}
 We have taken into account that
 LMA oscillations suppress both rates by about
 $0.55\pm0.02$ --- a value negligibly different from the
 low-energy limit of $P(\nu_e\to\nu_e)$ of eq.~(\ref{eq:PeeLMA}).

 In order to show the accuracy of our simplified analysis, we now compare its results
with the `exact' results of a solar-model-independent global analyses of solar and KamLAND data, 
performed as described in~\cite{BS}.
To perform such comparison we need to consider a well defined, simple and relevant sub-case:
we assume a non standard Beryllium flux but a standard CNO flux,
$\Phi_{\rm CNO}\approx 0.6\cdot 10^{9}/\cm^2{\rm s}$.
As previously discussed, this 
choice is motivated by the fact that
Beryllium neutrinos are more important,
that {\sc Borexino} should study them, and that
according to solar models 
the CNO contribution to any measured solar neutrino rates
is smaller than its experimental error.
In this way eq.\eq{CNOBe} reduces to
$\Phi_{^7{\rm Be}}=(4.9\pm 1.1)\cdot 10^{9}/\cm^2{\rm s}$.
This can be directly compared with the result of the global analysis, shown in  fig.\fig{BeB2005},
that can be approximatively summarized as
\begin{eqnsystem}{sys:exp}
\Phi_{^8\rm  B} &=& (5.0\pm0.2)\cdot
10^{6}/\cm^2{\rm s},\\
\Phi_{^7\rm Be}&=& (4.1\pm1.1)\cdot
10^{9}/\cm^2{\rm s}.\label{eq:Be}
\end{eqnsystem}
The global fit also includes the Chlorine rate, which
has a central value
about $2\sigma$ lower than the LMA prediction,
and consequently
somewhat reduces $\Phi_{^7\rm Be}$.
{}From these numbers, we see that the 
determinations of the Boron and Beryllium fluxes
that follow from our simplified 
analysis are quite adequate.
{\sc Borexino} will significantly improve over the present determination of $\Phi_{^7{\rm Be}}$.

\medskip


\medskip

The simplified analysis also determines our present knowledge of the  CNO flux.
As previously discussed solar neutrino experiments basically measured the linear combination
of CNO and Beryllium fluxes of eq.\eq{CNOBe}, that directly implies
an  upper bound on the CNO flux:
\begin{equation}\label{eq:CNO}
\Phi_{\rm CNO} < 6\cdot 10^9/\cm^2{\rm s}\qquad\hbox{ at $3\sigma$ (1 dof),}\end{equation}
which is one order of magnitude above solar model predictions.
Eq.\eq{CNO} updates the value first obtained in~\cite{BS} (section 4).
This constraint was also re-derived in~\cite{CNO}, emphasizing
that the result of eq.\eq{CNOBe} or of eq.\eq{CNO}
proves that the CNO cycle does not give the dominant contribution 
to the total solar luminosity $L_\odot$.
Indeed by converting  the neutrino flux $\Phi_{\rm CNO}$ into
the corresponding energy flux $L_{\rm CNO}$, eq.\eq{CNO} reads
$L_{\rm CNO} \circa{<} 0.1~L_\odot$.

\section{Oscillations with atmospheric frequency\label{sec:atm}}
Present data do not precisely determine  $|\Delta m^2_{23}|$ nor $\theta_{23}$
(see table~\ref{tab1} or fig.~\ref{fig:panta}), give an upper bound on 
$\theta_{13}$ and do not determine the sign of $\Delta m^2_{23}$ (i.e.\ if neutrinos have `normal' or `inverted' mass hierarchy).
Several experimental programs 
using long-baseline 
and reactor neutrinos
plan to to confirm the SK evidence 
and to improve on $\Delta m^2_{23}$, $\theta_{23}$ and $\theta_{13}$,
possibly measuring a non-zero value of the latter angle. From 
the point of view of the 3 neutrino framework these 
experimental programs seem adequately complete~\cite{LBL}, 
thanks  in particular to JHF~\cite{jhf}, possibly complemented with 
a new reactor neutrino experiment \cite{hub}.
Here, we discuss the present determination of atmospheric parameters 
and other implications of the existing data.

\paragraph{The global fit.}
The atmospheric 
 fit includes the final SKI data~\cite{SKI}, the final  {\sc Macro} data~\cite{Macro} and 
the latest K2K~\cite{K2K} data.
As in other similar analyses, the SK and K2K fits are
 extracted from the latest SK and K2K papers by 
`graphical reduction' (i.e.\ using a scale and a pencil) because
this procedure guarantees a more accurate result 
than an independent reanalysis.
We  employ the `standard' analysis of  final SKI data.
The SK collaboration also performs an alternative analysis,
by selecting the data which have the best resolution in $L/E_\nu$,
obtaining similar central values and somewhat different uncertainties.
(The CPT-violating fit of fig.\fig{CPT2005}a 
is instead based on an independent reanalysis of atmospheric data,
so its CPT-conserving limit can slightly differ from the CPT-conserving fit
of fig.\fig{panta}).

\paragraph{The parameter $\theta_{23}$.}
The value of this parameter 
can be obtained from a simple physics argument.
It is dominantly determined by SK multi-GeV $\mu$-like events
as $\sin^2 2\theta_{23}\simeq 2(1-
N_\uparrow/N_\downarrow )= 1.02\pm0.08$, where
$N_\downarrow\approx 400$ ($N_\uparrow\approx 200$) are the number of
down-ward (up-ward)  going $\mu$ events, that experience
roughly no oscillation (averaged oscillations).
A detailed analysis is needed to include the rest of the SK, K2K, {\sc Macro}
data, which however only mildly improve on 
the determination of $\sin^2 2\theta_{23}$.
An improved measurement of $\sin^22\theta_{23}$ from the up/down ratio of
atmospheric neutrinos could be performed
at a future Mton-scale atmospheric detector.\footnote{ 
One should take into account the few $\%$
effects of solar oscillations and of 
$\theta_{13}$ (see~\cite{peres} for a recent discussion)
which will be respectively better measured 
by solar and reactor/long-baseline experiments.}
Future long-baseline experiments with very intense conventional neutrino beams \cite{jhf}
will also lead to progress.

\paragraph{The parameter $|\Delta m^2_{23}|$.}
The present situation concerning $|\Delta m^2_{23}|$ is quite different:
SK cannot precisely measure it and cannot see a clear oscillation dip,
and a detailed analysis
is necessary to extract its central value and error.
Long-baseline experiments should significantly reduce 
the uncertainty on $|\Delta m^2_{23}|$
by identifying the energy at which neutrinos experience the first
oscillation dip.
This measurement has been already performed by K2K, 
but with poor statistics:  K2K achieves~\cite{K2K}
a determination of $|\Delta m^2_{23}|$ with 
central value and error
close to the one of SK~\cite{SKI}.

\paragraph{The parameter $\theta_{13}$.}
The CHOOZ constraint on $\theta_{13}$
is strongly correlated with the determination of $|\Delta m^2_{23}|$. As discussed in section~\ref{sec:PeeLow} 
solar data have a subdominant impact
on the determination of $\theta_{13}$,
comparable to the effect of changing the kind of analysis of SK 
atmospheric data (we used the `standard' SK analysis).
The statistically insignificant hint for a $\theta_{13}>0$ in fig.\fig{panta}
is mainly due to a small deficit of events in CHOOZ data
at lowest energies.

\paragraph{Other effects?}
Data show no significant hint for 
new effects beyond three massive neutrinos.
For example fig.\fig{CPT2005}a shows a global 
fit performed without assuming
that neutrinos and anti-neutrinos have the same
atmospheric mass splitting and mixing angle.
We see that the best-fit lies close to the CPT-conserving limit,
and that the atmospheric mass splitting in anti-neutrinos
is poorly determined.
Nevertheless, this is enough to strongly disfavor a CPT-violating interpretation 
of the LSND anomaly~\cite{CPT}.
Near-future long-baseline experiments will probably 
study only $\nu$ rather than $\bar\nu$.

\section{Non-oscillation experiments\label{sec:bb}}

In this section we discuss non-oscillation experiments  and consider 
the 3 non-oscillation parameters mentioned in the introduction.
Making reference to experimental sensitivities, 
the 3 probes should be ordered as follows: 
cosmology, $0\nu2\beta$ and finally $\beta$ decay.
Ordering them according to reliability would presumably result into 
the reverse list: cosmological results are based on untested assumptions, 
and $0\nu2\beta$ suffers from severe 
uncertainties in the nuclear matrix elements.
Even more, there is an interesting claim that the $0\nu2\beta$
transition has been detected~\cite{Klapdor}
(see section~\ref{sec:0nu2beta} for some remarks), there is a persisting 
anomaly in TROITSK $\beta$ decay,  and even in cosmology, there is 
one (weak) claim for  a positive effect. 
None of these hints can be considered as a  discovery of neutrino masses.
Several existing or planned experiments  will lead to progress in a few years. 

In this section, we assume three massive Majorana 
neutrinos and study the ranges of neutrino mass signals
expected on the basis of oscillation data,
updating and extending the results of~\cite{Feru}.
Our inferences are summarized in table~\ref{tab1} and obtained by marginalizing 
the full joint probability distribution for the oscillation parameters,
using the latest results discussed in the previous sections.

\begin{figure}
$$\includegraphics[width=7cm]{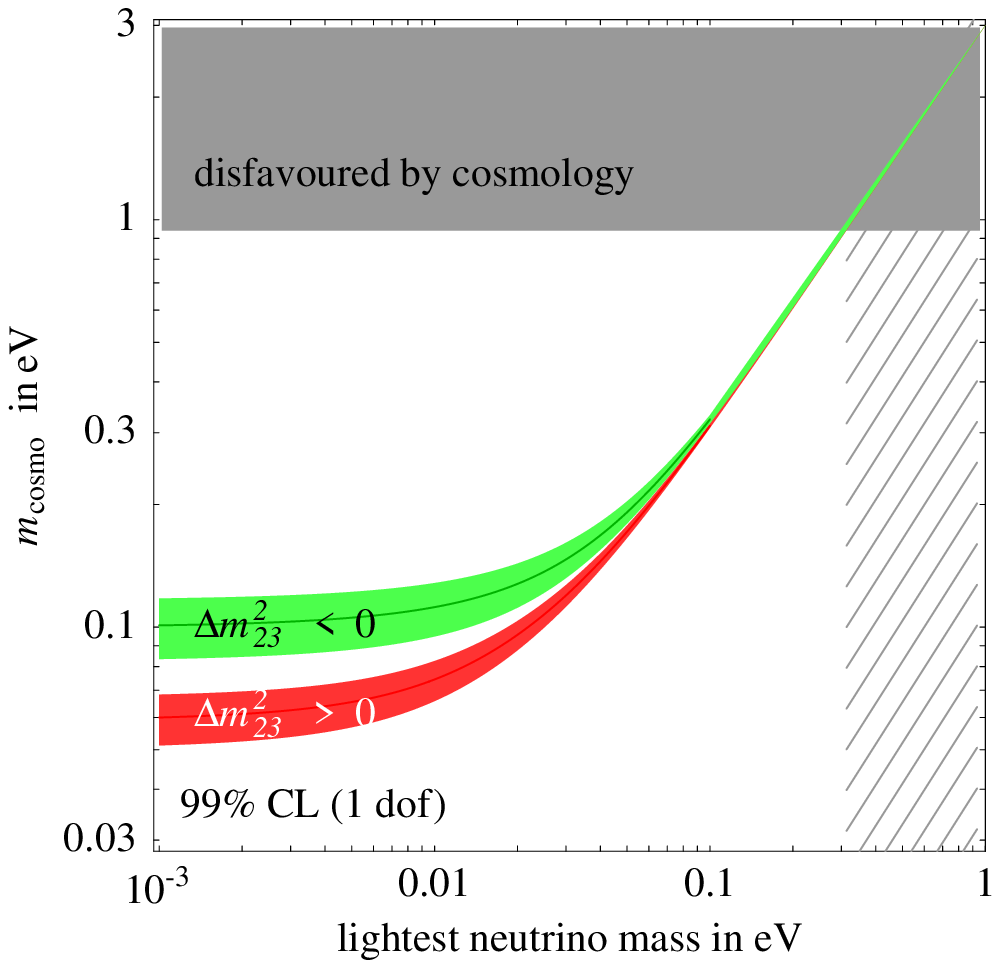}\qquad\qquad
\includegraphics[width=7cm]{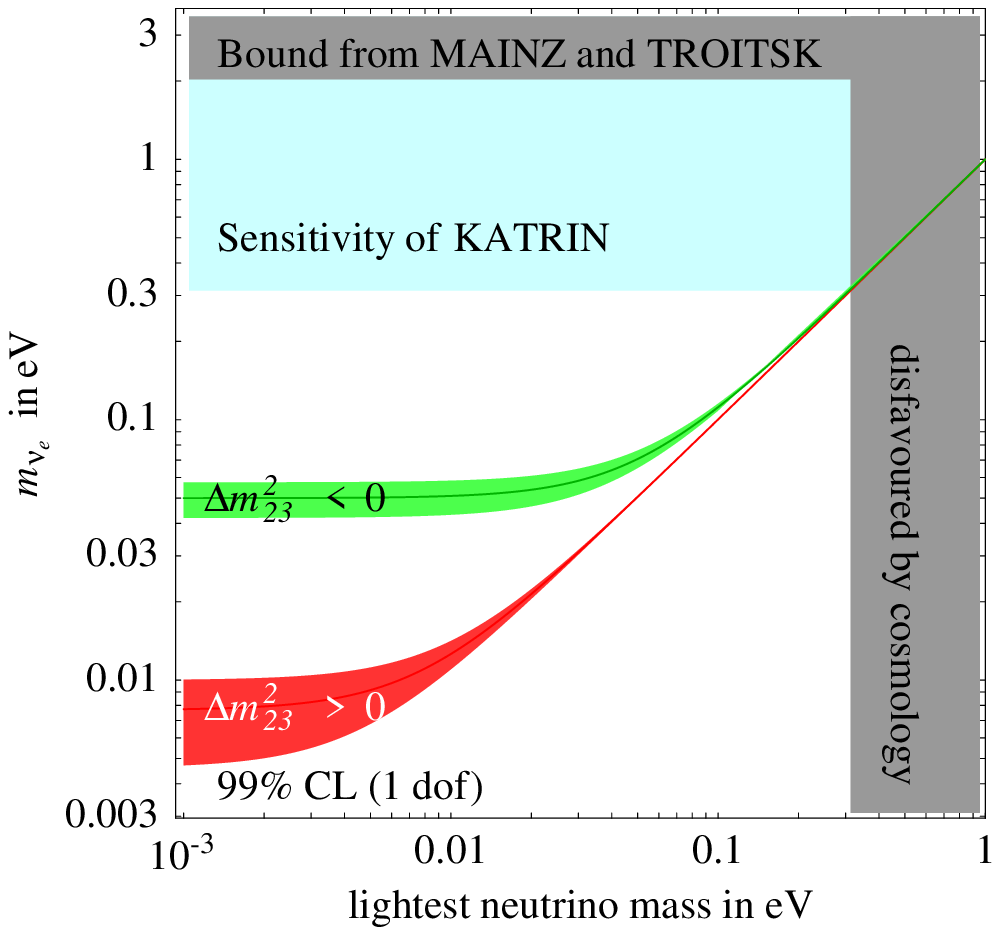}$$
\caption[x]{\label{fig:cosmobeta}\em $99\%$ CL expected ranges for the parameters $m_{\rm cosmo}=m_1+m_2+m_3$ 
probed by cosmology (fig.\fig{cosmobeta}a)
and 
$m_{\nu_e}\equiv (m\cdot m^\dagger)^{1/2}_{ee}$
probed by  $\beta$-decay (fig.\fig{cosmobeta}b)
as function of the lightest neutrino mass.
The darker lines show how the ranges would shrink if
the present best-fit values of oscillation parameters were 
confirmed with negligible error.}
\end{figure}

\subsection{Cosmology}
Cosmological observations are mostly sensitive to the
sum of neutrino masses:
$ m_{\rm cosmo}=m_1+m_2+m_3$,
that according to standard cosmology
controls the energy fraction $\Omega_\nu$ in non relativistic neutrinos
as $\Omega_\nu h^2 = m_{\rm cosmo}/ 93.5\eV$ where as usual
$h$ parameterizes the present value of the Hubble constant as
$h\equiv H_{\rm today}/(100 \hbox{km/s\,Mpc})$.
Cosmology does not distinguish Majorana from Dirac neutrino masses.

In order to convert CMB and LSS data into a constraint on neutrino masses
one needs to assume a cosmological model.
The ``WMAP constraint''~\cite{WMAP} assumes
that the observed structures are generated by Gaussian adiabatic primordial scalar fluctuations
with flat spectral index $n$
evolved in presence of the known SM particles, of cold dark matter and of a cosmological constant.
This standard model of cosmology seems consistent with all observations.
Furthermore, it is assumed that observed luminous matter tracks the dark matter density 
up to a scale-independent bias factor.
The data-set includes Lyman-$\alpha$ data about structures
on scales so small that perturbations are no longer a minor correction
to a uniform background.
These data are sensitive to neutrino masses (and thereby somewhat affect the global
cosmological fit) but
might be affected by non linear evolution effects, which are not fully understood.
In summary, cosmology presently gives the dominant constraint,
which however rests on untested assumptions and on risky systematics.

In the future the sensitivity to neutrino oscillations will improve thanks to better CMB data
and to new LSS measurements less plagued by potential systematic effects.
If cosmology were simple
(e.g.\ a spectral index $n=1$, no tensor fluctuations,\ldots)
then it seems possible to detect even neutrino masses 
as small as allowed by oscillation data~\cite{CosmoFuture}.
The expected ranges of $m_{\rm cosmo}$ are reported in the lowest row of table~\ref{tab1}
in the limiting case where the lightest neutrino is massless,
and in fig.\fig{cosmobeta}a in the general case.

\subsection{Direct search via $\beta$ decay}
 $\beta$-decay experiments dominantly
 probe the quantity $m_{\nu_e}\equiv (m\cdot m^\dagger)_{ee}^{1/2}$.
 If neutrinos are quasi-degenerate, $m_{\nu_e}$ is their common mass.
The MAINZ and TROITSK experiments obtained 
comparable limits:
$m_{\nu_e}^2=-1.2\pm 2.2 \pm 2.1$~eV$^2$~\cite{MAINZ} and
$m_{\nu_e}^2=-2.3\pm 2.5 \pm 2.0$~eV$^2$~\cite{TROITSK} respectively.
The 95\% bounds quoted by the experimental collaborations
agree with the values obtained in Gaussian approximation. Thus, 
we combine the two measurements by summing errors in quadrature and get
\begin{equation} m_{\nu_e}^2=-1.7\pm 2.2 \mbox{ eV}^2\qquad\hbox{i.e.}\qquad
m_{\nu_e}<2.0\,\eV\hbox{ at 99 \% C.L.}\end{equation}
In order to study what oscillation data imply on the value of $m_{\nu_e}$ we
write it in terms
of the neutrino masses $m_i$ and of the mixing angles $\theta_{ij}$ as
\begin{equation}\label{eq:mnue}
m_{\nu_e} =
\bigg[ \cos^2 \theta_{13}(m_1^2 \cos^2\theta_{12} +
m_2^2  \sin^2\theta_{12}) + m_3^2 \sin^2\theta_{13}\bigg]^{1/2}.
\end{equation}
In the case of normal mass hierarchy, $m_1\ll m_2\ll m_3$
oscillation data imply the 99\% CL range
$m_{\nu_e} = (4.6\div 10)\meV$.
In the case of inverted hierarchy, $m_3\ll m_1<m_2$ one gets
$m_{\nu_e} = (42\div 57)\meV$.
The last number if a factor 5 below the planned sensitivity of {\sc Katrin}~\cite{Katrin}.

It is immediate to obtain the ranges corresponding to the generic case of
a non vanishing lightest neutrino mass $m_{\rm lightest}$
(where $m_{\rm lightest}=m_1$ in the case of normal hierarchy
and $m_{\rm lightest}=m_3$ in the case of inverted hierarchy).
As clear from the definition $m_{\nu_e}^2\equiv (m\cdot m^\dagger)_{ee}$ or from
the more explicit expression in eq.\eq{mnue} one just needs to add $m_{\rm lightest}^2$ 
to $m_{\nu_e}^2$.  The resulting bands at $99\%$ C.L.\  are plotted in fig.\fig{cosmobeta}b.

\subsection{Neutrino-less double-beta decay}\label{sec:0nu2beta}
Updating the results of~\cite{Feru},
in the 3 neutrino framework we discuss the connection 
with oscillations, the bound from $0\nu2\beta$ on neutrino masses
and the possible  hint of a signal.

\paragraph{IGEX, C{\small UORICINO} and nuclear uncertainties.}
First, we recall that recently two experiments produced new relevant data.
The first is IGEX~\cite{ig1}, now terminated, which used 86\% 
enriched ${}^{76}$Ge (with exposure $7\cdot 10^{25}$ nuclei$\cdot$yr).
The second is {\sc Cuoricino}~\cite{cuo2}, 
now running, which uses 34\% natural ${}^{130}$Te
(with exposure $1.4\cdot 10^{24}$ nuclei$\cdot$yr, and 
the plan to collect 10 times more data in about 6 years). 
Both experiments reached a good level of background
(about $0.18/\rm  keV\cdot kg\cdot yr$),
good detection  efficiencies of 
70\% and 84\%,  and respectable energy resolutions,
the full width at half maximum being 
4~keV and 7~keV respectively. 
For comparison, the Heidelberg--Moscow (HM) data-set 
with pulse-shape-discrimination \cite{HM}  
has exposure 
$2.5\cdot 10^{26}$ nuclei$\cdot$yr and background
$0.06/\rm ( keV\cdot kg\cdot yr)$.

As in~\cite{Feru} we quote the limits on $|m_{ee}|$ adopting the $0\nu2\beta$ nuclear
matrix elements ${\cal M}_0$ computed in \cite{staudt90}.
To use a different calculation with matrix element ${\cal M}$, 
just rescale by the factor $h={\cal M}_0/{\cal M}$,
which depends on the nucleus studied.\footnote{The
symbol $h$ has different meaning for cosmology and for $0\nu 2\beta$,
as should be clear from the context.}
We always explicit the factors $h$ when quoting an experimental result on $0\nu 2\beta$.
We prefer to show such uncertainty explicitly 
rather than attempting to evaluate the 
theoretical error on matrix elements.\footnote{It would be useful if future calculations of matrix elements
could provide error estimates.}
Different published computations find $h$ in the following ranges
\begin{equation}0.3~\cite{staudt92}< h({}^{76}{\rm Ge}) < 2.4~\cite{caurier96},\qquad
0.4~\cite{staudt92}  <h({}^{130}{\rm Te})< 2.7~\cite{rodin03}.\end{equation}
When comparing two results on
$|m_{ee}|$ obtained with different nuclei one needs to
consider the ratios of $h$, e.g., $h(^{130}\mbox{Te})/h(^{76}\mbox{Ge})$. 
This quantity is also uncertain, spanning the following range:
\begin{equation}\label{eq:h/h}
0.3~\cite{engel88} < h(^{130}\mbox{Te})/h(^{76}\mbox{Ge})< 1.7~\cite{rodin03}. 
\end{equation}

\begin{table}
\begin{center}
\begin{tabular}{| rl||c|c|c|c|c |}
\hline
\multicolumn{2}{|c||}{Nucleus and} & observed & background  & expected & 99\% C.L.\ bound \\ 
\multicolumn{2}{|c||}{experiment} & events, $n$  & events, $b$    & signal & on  $|m_{ee}|/h$  \\ \hline\hline
${}^{76}$Ge& HM~\cite{HM}  & 21& $20.4\pm 1.6$  &  ~76 $\,|m_{ee}/\eV|^2/h^2$     &  0.44 eV \\
${}^{76}$Ge&IGEX~\cite{ig1,ig2} & 9.6& $17.2\pm 2\phantom{.0}$    &  23.5$|m_{ee}/\eV|^2/h^2$    &  0.55 eV\\
${}^{130}$Te &{\sc Cuoricino}~\cite{cuo1,cuo2} &24& $35.2\pm 4\phantom{.0}$ & 21.5$|m_{ee}/\eV|^2/h^2$&  0.62 eV\\ \hline
\end{tabular}
\end{center}
\caption{\em 
Numbers of observed events, expected background 
and predicted signal in the most sensitive $0\nu2\beta$ experiments.
The last column shows the $99\%$ C.L.\ constraint on $|m_{ee}|/h$,
where $h\sim 1$ parameterizes the uncertain $0\nu2\beta$ nuclear matrix element and
depends on the nucleus studied.
\label{bbb}}
\end{table}

Now we convert $0\nu 2\beta$ data into a constraint on $|m_{ee}|$
following the same  simple procedure employed by the HM collaboration~\cite{HM}:
the Poisson likelihood  of having $s$ signal events is
${\cal L}(s)\propto e^{-s} (b+s)^n$, where $n$ and $b$  are the 
numbers of observed and expected background events 
in the 3$\sigma_E$ window around the $Q$ value of the $0\nu 2\beta$,
and $\sigma_E$ is the energy resolution of the apparatus.
We therefore evaluate $n$ and $b$ 
in the 10 keV region around $Q=2039$~keV for IGEX~\cite{ig1,ig2} and 
in the 18 keV region around $Q=2529$~keV for {\sc Cuoricino}~\cite{cuo1,cuo2}.
Results are collected in table~\ref{bbb}.  
For both experiments the observed 
number of events is slightly below the expected background. 
In this paper we systematically employ the Gaussian approximation:
the $99\%$ constraint (2.58$\sigma$) on $|m_{ee}|$ is computed by defining
$\chi^2\equiv -2\ln{\cal L}$, marginalizing it with respect to the uncertainty in the background,
and finally quoting the value of $|m_{ee}|$ such that
$\chi^2(m_{ee})-\chi^2(m_{ee}=0) = 2.58^2$.
Different approaches to statistical inference give slightly different constraints.

Since HM and IGEX are both based on 
${}^{76}$Ge, it is possible to obtain a more stringent  combined bound,
$|m_{ee}|<0.38\ h$ eV.\footnote{We cannot combine HM/IGEX with 
{\sc Cuoricino} in a reliable manner, 
due to the uncertainty on the relative nuclear matrix element discussed above.
Using the matrix elements of \cite{staudt90} for ${}^{76}$Ge {\em and} ${}^{130}$Te, 
would give $|m_{ee}|<0.34\ h$ eV at $99\%$ C.L.\ with $h=1$.}

\begin{figure}
$$\hspace{-8mm}
\includegraphics[width=8.5cm]{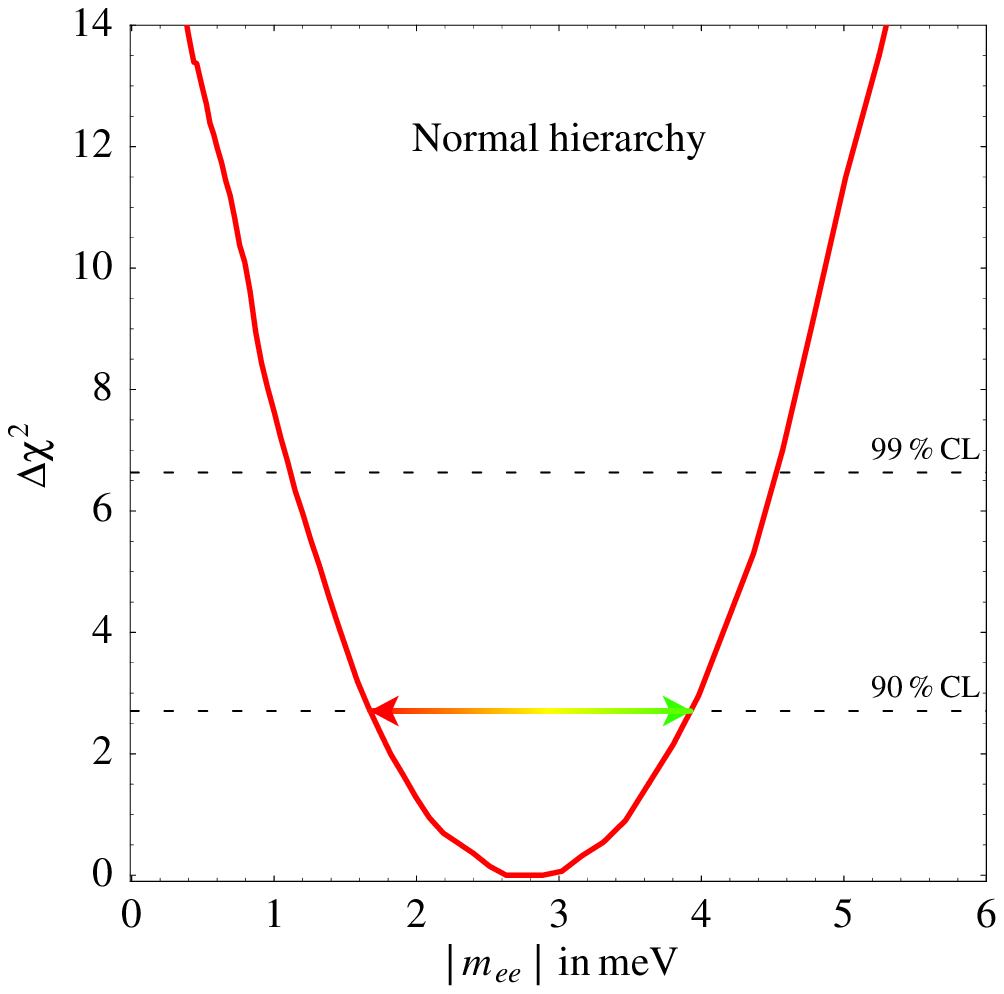}\hspace{1cm}\includegraphics[width=8.5cm]{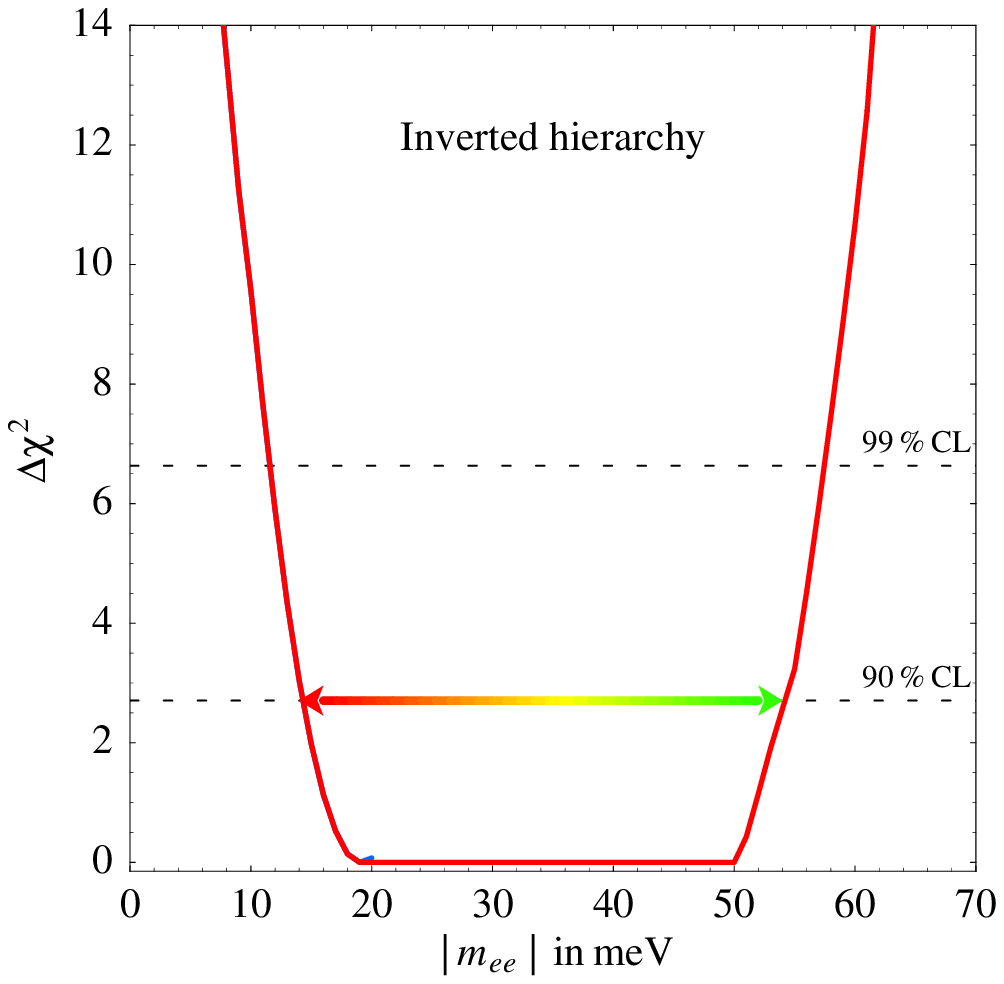}$$
$$\hspace{-8mm}
\includegraphics[width=8.5cm]{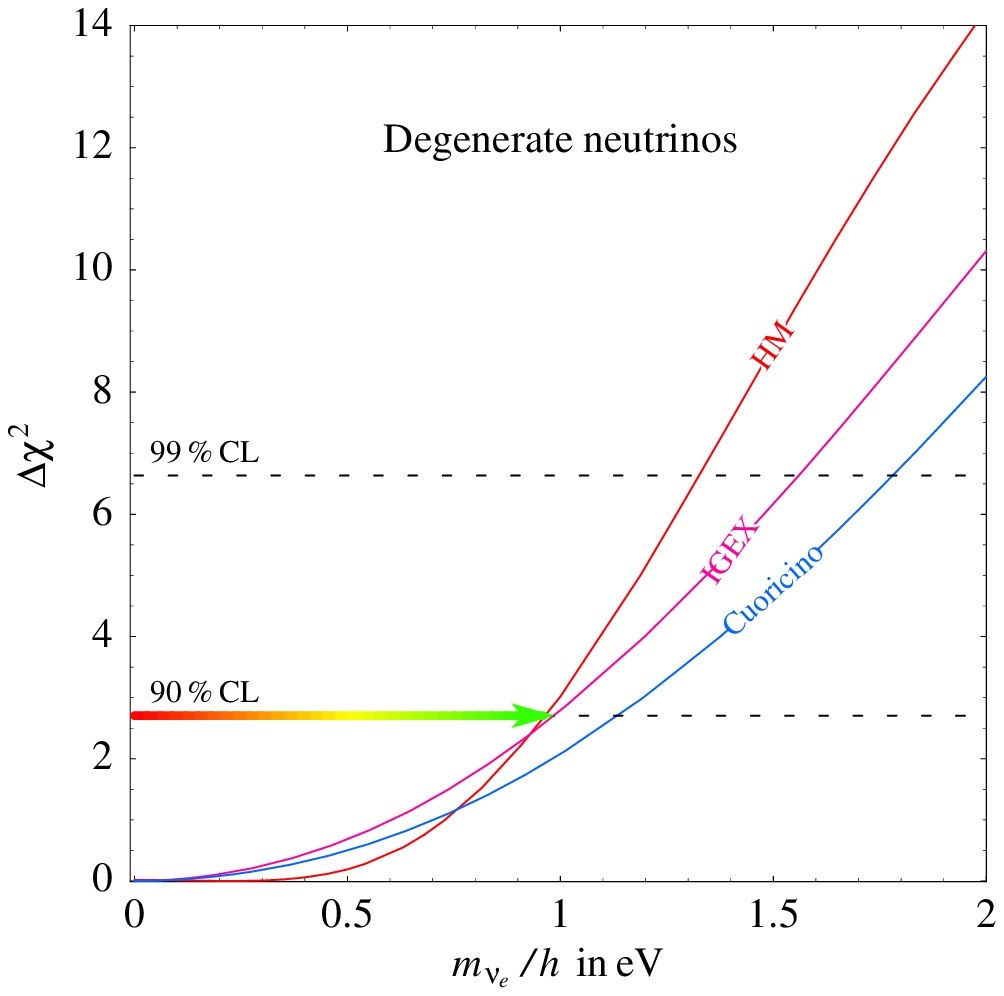}\hspace{1cm}\includegraphics[width=8.5cm]{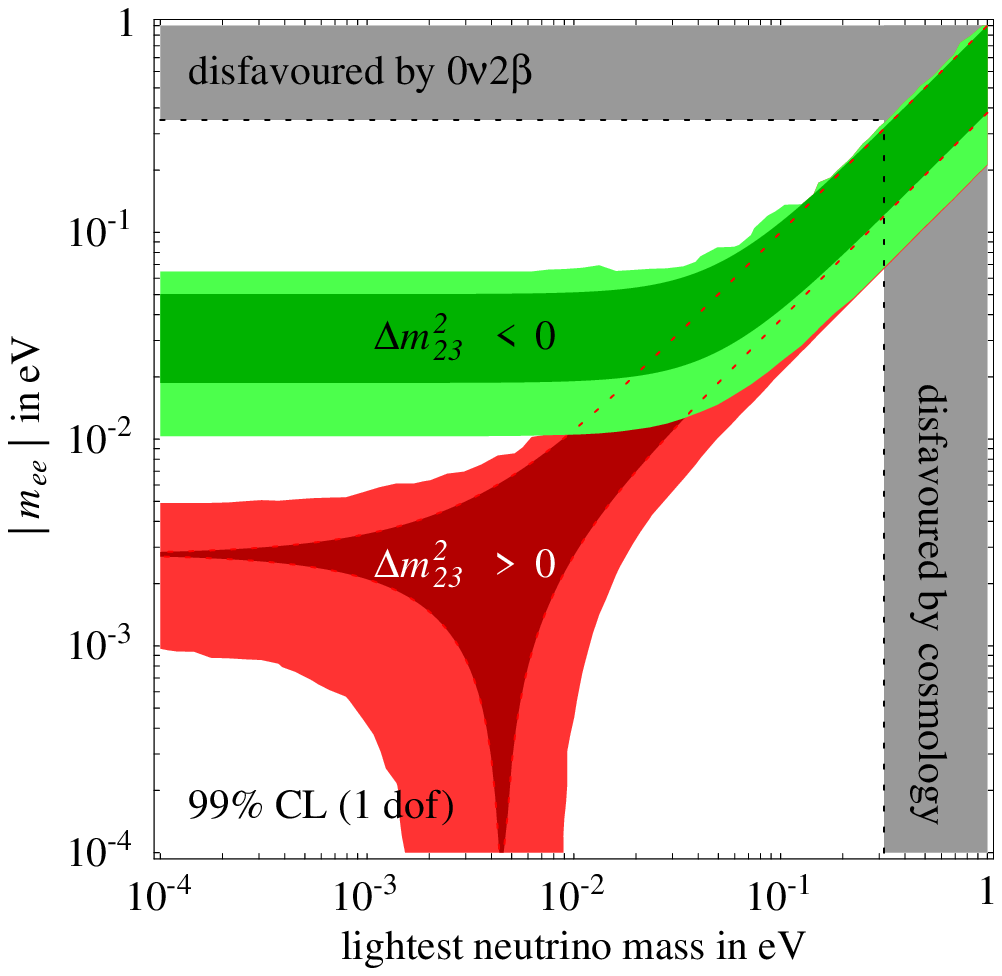}$$
\caption{\label{fig:postsalt}\em Predictions for $|m_{ee}|$ assuming a 
hierarchical 
(fig.\fig{postsalt}a) and
inverted (fig.\fig{postsalt}b)  neutrino spectrum. 
In fig.\fig{postsalt}c we update 
the upper bound on the mass of quasi-degenerate
neutrinos implied by $0\nu2\beta$ searches.
The factor $h\approx 1$ parameterizes the uncertainty in the 
nuclear matrix element (see sect.\ 2.1).
In fig.\fig{postsalt}d we plot the $99\%$ CL range for $m_{ee}$
as function of the lightest neutrino mass,
thereby covering all spectra.
The darker regions show how the $m_{ee}$ range would shrink if
the present best-fit values of oscillation parameters were 
confirmed with negligible error.
}\end{figure}

\paragraph{Inference on $|\mb{m}_{ee}|$ from oscillations.}
Using the latest oscillation data, we study the expected range of $|m_{ee}|$.
We follow the procedure described in~\cite{Feru}.
Assuming three CPT-invariant massive Majorana neutrinos we get:
\begin{itemize}

\item[a)]  In the case of {\bf normal hierarchy} (i.e.\ $m_1\ll m_2\ll m_3$, or $\Delta m^2_{23}>0$)
the $ee$ element of the neutrino mass matrix probed by
$0\nu2\beta$ decay experiments
can be written as $|m_{ee}| = |e^{2i \alpha}  m_{ee}^{\rm sun} + e^{2i\beta} m_{ee}^{\rm atm}|$
where $\alpha,\beta$ are unknown Majorana phases and 
the `solar' and `atmospheric' contributions can be predicted from oscillation data.
The solar contribution to $m_{ee}$
is $m_{ee}^{\rm sun} = (\Delta m^2_{12})^{1/2}\sin^2\theta_{12}\cos^2\theta_{13}=
( 2.78\pm 0.22)~\,\hbox{meV}$.
The bound on $\theta_{13}$ implies $m_{ee}^{\rm atm} =(\Delta m^2_{23})^{1/2}\sin^2\theta_{13}
< 1.7\,\hbox{meV}$ at 99\% CL.
By combining these two contributions we get
the range reported in table~\ref{tab1}.
The precise prediction is shown in fig.\fig{postsalt}a.
Unlike in our previous results~\cite{Feru}, 
a non zero contribution is now guaranteed
at high confidence level, because data now tell that the `solar' contribution is larger
than the `atmospheric' contribution, so that a cancellation is not possible.

\item[b)] In the case of {\bf inverted hierarchy} 
(i.e.\ $m_3\ll m_1\approx m_2$, or $\Delta m^2_{23}<0$), from
the prediction
$|m_{ee}| \approx |\Delta m^2_{23}|^{1/2}\times
|\cos^2\theta_{12} +  e^{2i\alpha} \sin^2 \theta_{12}|\cos^2 \theta_{13}$
we get the range reported in table~\ref{tab1}
(see also~\cite{MuraPenya}).
The precise prediction is shown in fig.\fig{postsalt}b.
The main uncertainty on $|m_{ee}|$ is due to the Majorana phase $\alpha$,
rather than to the oscillation  parameters.
If the present central values were confirmed with infinite precision we would still have the loose range
$|m_{ee}|=(19\div 50) \meV$ at any C.L.

\end{itemize}
The $|m_{ee}|$ ranges for normal and inverted hierarchy do not overlap. 
Values of $|m_{ee}|$ outside these ranges are possible if the 
lightest neutrino mass
is not negligible, as shown in fig.\fig{postsalt}d.
In the inverted hierarchy case ($\Delta m^2_{23}<0$)  a non zero lightest neutrino mass 
can only make $|m_{ee}|$ larger than in case b).
The darker regions in fig.\fig{postsalt}d 
show how the predicted range of $|m_{ee}|$ would shrink 
if the present best-fit values of
oscillation  parameters were confirmed with infinite precision.
The two funnels present for $\Delta m^2_{23}>0$
have an infinitesimal width because we assume $\theta_{13} = 0$
and would have a finite width if $\theta_{13}\neq 0$.
\begin{itemize}

\item[c)]
The case of quasi-{\bf{}degenerate neutrinos} with mass $m_{\nu_e}$
corresponds to the upper region 
of  fig.\fig{postsalt}d.  $|m_{ee}|$ and $m_{\rm cosmo}$ are related to $m_{\nu_e}$  as
\begin{equation}
\label{eq:neqh}m_{\rm cosmo}=3 m_{\nu_e}\qquad\hbox{and}\qquad
0.24\,m_{\nu_e}<|m_{ee}|<m_{\nu_e} \hbox{ at $99\%$ C.L.}
\end{equation}
The lower bound  on $|m_{ee}|$ holds
thanks to the fact that solar data exclude a maximal solar mixing
and that CHOOZ requires a small $\theta_{13}$.
\end{itemize}

\paragraph{Upper bound on neutrino masses from $0\nu 2\beta$.}
The above result means that by combining oscillation data with
the $0\nu2\beta$ upper bound on $|m_{ee}|$
implies an upper bound on the parameter $m_{\nu_e}$ probed by $\beta$ decay experiments~\cite{Feru}.
In view of present values, such $m_{\nu_e}$ corresponds
to the common mass of quasi-degenerate neutrinos.
This bound is shown in fig.\fig{postsalt}c, and 
depends on the $0\nu2\beta$ nuclear matrix elements,
parameterized by the uncertain  factors $h\sim 1$.
Therefore the results of {\sc Cuoricino}, obtained
with a nucleus different than HM and IGEX, add confidence
in the  result.
For $h=1$ the combined constraint is $m_{\nu_e}/h<1.0\,\eV$ at $99\%$ C.L.
This constraint is stronger than the $\beta$-decay constraint 
(but holds under the additional assumption that neutrinos are Majorana particles)
and is weaker than the cosmological constraint
(but  needs no assumptions about cosmology).

In all above discussions we employed the bound on $|m_{ee}|$ from
HM data, as published by the HM collaboration~\cite{HM}.

%



\begin{figure}
$$\hspace{-4mm}\includegraphics[height=5cm]{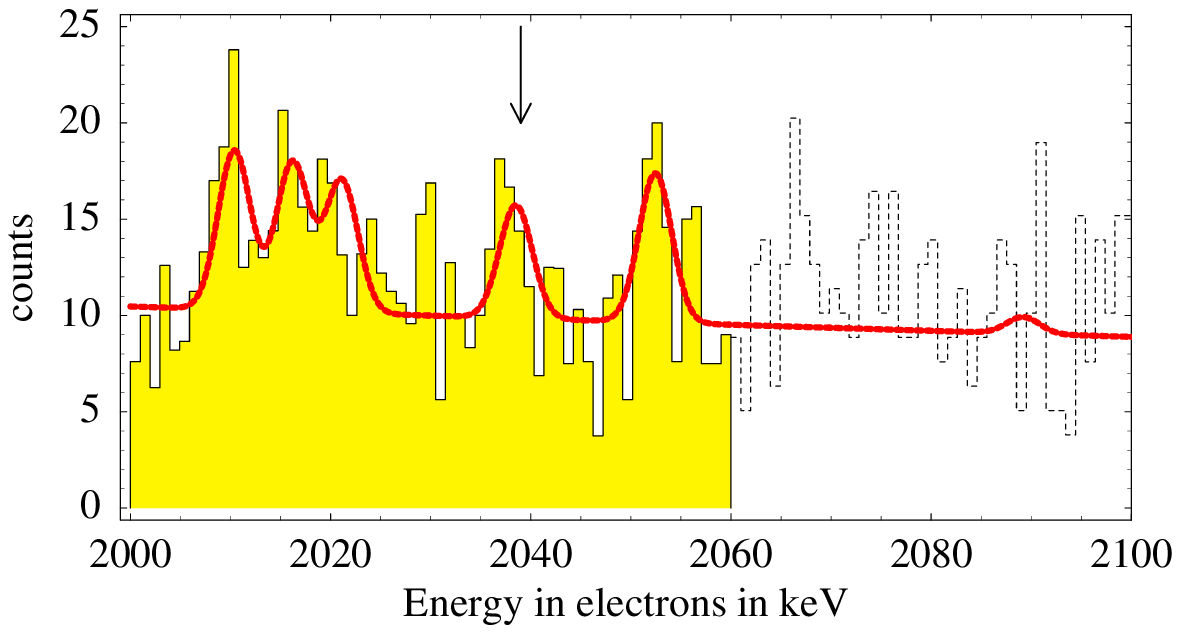}\includegraphics[height=5cm]{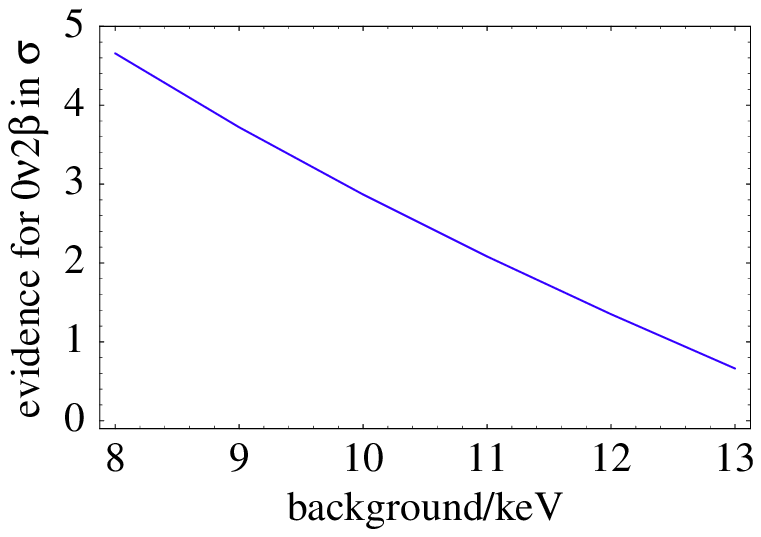}$$
\caption{\label{fig:Klapdor2004}\em Fig.\fig{Klapdor2004}a: the latest HM data~\cite{KlapdorLast}
($71.7{\rm kg}\cdot{\rm yr}$)
used to claim a $4.2\sigma$ evidence for $0\nu2\beta$.
Fig.\fig{Klapdor2004}b: the statistical significance of the $0\nu2\beta$ signal, as
function of the assumed flat component of the background.}
\end{figure}

\paragraph{Remarks on the hint for $0\nu 2\beta$.}
While the HM collaboration used their data  to set a bound on $|m_{ee}|$, 
some members of the HM collaboration reinterpreted the data 
as an evidence for $0\nu2\beta$~\cite{Klapdor}.
According to this claim, the latest HM data~\cite{KlapdorLast} plotted in 
fig.\fig{Klapdor2004}a
contain a $4.2\sigma$ evidence for 
a $0\nu2\beta$ peak (indicated by the arrow).
In these latest results the peak is more visible than in  
latest published HM data~\cite{HM},
partly thanks to higher statistics (increased from 53.9 to 71.7 kg yr)
and partly thanks to an `improved analysis'~\cite{KlapdorLast}).

This claim is controversial, mainly because
one needs to fully understand the background before being
confident that a signal has been seen. 
In order to allow a better focus on this key issue,
we present fig.\fig{Klapdor2004}b, which should be uncontroversial.
It shows  the statistical significance\footnote{Defined in Gaussian approximation
as $(\chi^2_{\rm no~signal}-\chi^2_{\rm best~fit})^{1/2}$ where
$\chi^2 =  -2\ln{\cal L}$ and ${\cal L}$ is the likelihood
computed combining statistical Poissonian uncertainties with other
systematic uncertainties.} of the $0\nu2\beta$ signal
as a function of the {\em true} background level $b$,
assumed to be quasi-flat close to the $Q$-value of $0\nu2\beta$,
$Q\approx 2039\,{\rm keV}$.
The crucial issue is: how large is $b$?
The HM collaboration earlier claimed \cite{HM}
$b=(13.6\pm0.7){\rm events}/(71.7\,{\rm kg}\,{\rm yr}\cdot{\rm keV})$.
In such a case the statistical significance of 
the signal would be less than $1\sigma$, see fig.\fig{Klapdor2004}b.
This can be considered as the upper bound on $b$ computed assuming that all
events in a wider range around $Q$ come from a quasi-flat background.

A statistically significant hint for $0\nu2\beta$ is obtained if one can show that $b$ is lower.
The continuous line in fig.\fig{Klapdor2004}a shows a fit of HM data
using a tentative model of the background~\cite{Klapdor},
assumed to have a quasi-flat component $b$
(mainly due to `natural' and `cosmogenic' radioactivity)
plus some peaks due to faint $\gamma$ lines of $^{214}$Bi, which is
a radioactive impurity present in the apparatus
(from the $^{238}$U decay chain).
Their positions and intensities can be estimated from tables of nuclear decays;
however they are modified by ${\cal O}(1)$ factors 
by detector-related effects which
depend on the unknown localization of $^{214}$Bi
(see~\cite{KlapdorLast} and ref.s therein).
The fit in fig.\fig{Klapdor2004}a is performed by 
allowing the intensity of each line to freely vary.
In this way part of the background is interpreted as $^{214}$Bi peaks,
thereby reducing the quasi-flat component $b$.
Proceeding in this way, we find that the statistical 
significance of the $0\nu2\beta$ signal is about $2.7\sigma$.
This is the best we can do with the data at our disposal.
This kind of analysis was proposed in~\cite{Feru} and 
has been adopted in~\cite{KlapdorLast}.

However, some details in its implementation prevent this analysis from fully
reaching its goal, which is determining $b$ from regions with no peaks.
1) The latest data have been published only below 2060 keV.
(Above 2060 keV in fig.\fig{Klapdor2004}a we plotted HM data
artificially rescaled to account for the larger statistics).
Below 2060 keV there are little energy ranges with no peaks.
Including the old data above 2060 keV in
the fit would reduce the significance of 
the signal down to about $2.2\sigma$.
2) HM data contain hints of extra unidentified 
spurious peaks at specific energies 
(at 2030 keV and above 2060 keV).
Fitting data assuming that these extra peaks can
be present at arbitrary energies with arbitrary intensities
reduces $b$ and enhances the statistical significance of the signal.

Various future experiments plan to test the claim of~\cite{Klapdor}.
The most direct test requires using the same technique 
(germanium detectors) but in the worst case, when the observed hint of
a signal is due to some irreducible (hypothetical) background,
a safe test requires a different type of detector.
As previously discussed {\sc Cuoricino} employs a different nucleus:
its capabilities relative to HM and IGEX depend on the uncertain relative
nuclear matrix element.
For instance, with the lowest value in eq.\eq{h/h}
{\sc Cuoricino} is already testing the claim
of \cite{Klapdor};  
for intermediate values new data of {\sc Cuoricino}
will significantly test the claim; 
for the highest value {\sc Cuoricino} will be not sufficient.

In our view, a discussion of what should be considered as a convincing evidence for $0\nu2\beta$,
is anyway useful or necessary,
because any  experiment (past and future) needs to confront with this issue.

\section{Conclusions}
Assuming oscillations of three active massive neutrinos we updated the
determination of the oscillations parameters, at the light of latest experimental data.
Results are shown in table~\ref{tab1} and in fig.\fig{panta}.
We notice that the parameters $\Delta m^2_{12},\theta_{12},\theta_{23}$
are now dominantly determined by simple and robust sub-sets of data,
such that simple arguments give the same final result as global analyses.
Pieces of data that play a sub-dominant r\^ole in parameter determination
allow to test our assumptions: e.g.\ allowing neutrinos and anti-neutrinos
to have different masses and mixings gives the CPT-violating fit of fig.\fig{CPT2005}.
Present data do not contain evidence for extra effects.

We also tried to study in a simple and general way related topics,
such as the determination of the limiting survival probability of solar neutrinos 
at small and large energies, 
the present knowledge of solar neutrino fluxes, etc.
Approximated general results have been compared with `exact' results of
 global fits performed in specific cases.
 
 Finally, updating the results of~\cite{Feru},
   we studied how oscillation data allow to infer the combination of neutrino masses
 probed by cosmology, $\beta$-decay and $0\nu2\beta$-decay experiments,
 and discussed the present experimental situation.

\footnotesize
\begin{multicols}{2}
 
\end{multicols}


\begin{thebibliography}{99}
 
 
\bibitem{Chlorinelast}
The results of the Homestake experiment are reported in
\art{B.T. Cleveland et al.}{Astrophys. J.}{496}{505}{1998}.


\bibitem{Galliumlast} 
Latest {\sc Gallex} and SAGE data have been presented in a talk by C. Cattadori
at the `Neutrino 2004' conference (Paris, 14-19 June), 
web site neutrino2004.in2p3.fr.

\bibitem{Gallex} \art{{\sc Gallex} collaboration}{\PL}{B447}{127}{1999}.


\bibitem{SAGE}
SAGE Collaboration,  J.\ Exp.\ Theor.\ Phys.\  {95} (2002) 181  [astro-ph/0204245].



\bibitem{SKlast}
\hepart[hep-ex/0205075]{Super-Kamiokande collaboration}. 

\bibitem{SNOlast} {SNO collaboration}, nucl-ex/0204008 and
nucl-ex/0204009.

\bibitem{SNOsaltfinal} \hepart[nucl-ex/0502021]{SNO collaboration}.


\bibitem{KL2004} \hepart[hep-ex/0406035]{KamLAND collaboration}.

\bibitem{SKI}
  \hepart[hep-ex/0501064]{SuperKamiokande collaboration}.
  
  \bibitem{Macro}
  \art[hep-ex/0304037]{MACRO collaboration}{\PL}{B566}{35}{2003}.

\bibitem{K2K} 
\hepart[hep-ex/0411038]{K2K collaboration}.


   \bibitem{Klapdor}
   H.~V.~Klapdor-Kleingrothaus, A.~Dietz, H.~L.~Harney and I.~V.~Krivosheina,
  Mod.\ Phys.\ Lett.\ A {16} (2001) 2409
  [hep-ph/0201231].

  

\bibitem{KlapdorLast}
H.~V.~Klapdor-Kleingrothaus, A.~Dietz, I.~V.~Krivosheina and O.~Chkvorets,
  Nucl.\ Instrum.\ Meth.\ A {522} (2004) 371
  [hep-ph/0403018].


\bibitem{LSND}
\art[hep-ex/0104049]{LSND collaboration}{\PR}{D64}{112007}{2001}.


\bibitem{which}  A.~Strumia and F.~Vissani,
  JHEP {0111} (2001) 048 [hep-ph/0109172].


\bibitem{BP}
\art[astro-ph/0010346]{J.N. Bahcall, S. Basu, M.H. Pinsonneault}{Astrophys. J.}{555}{990}{2001}.


\bibitem{LUNA} \hepart[nucl-ex/0312015]{LUNA collaboration}.



\bibitem{sunfitsalt}
\art[hep-ph/0102234]{P. Creminelli et al.}{J. HEP}{05}{2001}{052}.
Its e-print version, has been updated
including the recent data from SNO.



\bibitem{CPT} \hepart[hep-ph/0201134]{A. Strumia}.
The solar CPT-violating fit has been re-emphasized in
H.~Murayama,  Phys.\ Lett.\ B {597} (2004) 73 [hep-ph/0307127] and in
\hepart[hep-ph/0406301]{A. de Gouvea, C. Pe\~na-Garay}.
The atmospheric CPT-violating fit has been also studied by the SK collaboration,
see the talk by E. Kearns at the Neutrino 2004 conference (Paris, 14-19 June), 
web site neutrino2004.in2p3.fr.
See also  M.~C.~Gonzalez-Garcia, M.~Maltoni and T.~Schwetz,
  Phys.\ Rev.\ D {68} (2003) 053007.
  
  

\bibitem{BS}
A. Strumia and R. Barbieri, JHEP 12 (2000) 016.  
Its e-print version, hep-ph/0011307, contains updates and additional discussions
not present in the published version.


\bibitem{theta13sun}
  S.~Goswami, A.~Bandyopadhyay and S.~Choubey,
  Nucl.\ Phys.\ Proc.\ Suppl.\  {143} (2005) 121 [hep-ph/0409224].
B.~S.~Koranga, M.~Narayan and S.~Uma Sankar, hep-ph/0503092.


\bibitem{roadmaps} \art{J.N. Bahcall, C. Pe\~na-Garay}{JHEP}{004}{0311}{2003}.
\hepart[hep-ph/0410283]{A. Bandyopadhyay, S. Choubey, S. Goswami, S.T. Petcov}.
  


\bibitem{sterile}
For a recent extensive study see
M. Cirelli et al, Nucl. Phys. B708 (2005) 215 [hep-ph/0403158].
For short summaries see
 A.~Strumia,  Nucl.\ Phys.\ Proc.\ Suppl.\  {143} (2005) 144
  [hep-ph/0407132] and  M.~Cirelli, astro-ph/0410122.  
  
  
\bibitem{Borexinopep}
C. Galbiati, A. Pocar, D. Franco, A. Ianni, L. Cadonati, S. Schonert,  hep-ph/0411002.

\bibitem{CNO}
J.~N.~Bahcall, M.~C.~Gonzalez-Garcia and C.~Pena-Garay,
    Phys.\ Rev.\ Lett.\  {90} (2003) 131301.
      


\bibitem{LBL} For a recent summary of capabilities of 
neutrino beam experiments see M. Lindner, hep-ph/0503101.
If $\theta_{13}=0$ it seems practically impossible to discriminate
the kind of neutrino mass hierarchy with oscillation experiments,
as recently discussed in 
A.~de Gouvea, J.~Jenkins and B.~Kayser, hep-ph/0503079.



\bibitem{jhf}
  \hepart[{\rm  ``The JHF-Kamioka neutrino project''}, 
hep-ex/0106019]{Y.~Itow {\it et al.}}


\bibitem{hub}
  P.~Huber, M.~Lindner, T.~Schwetz and W.~Winter,
  Nucl.\ Phys.\ B {665} (2003) 487
  [hep-ph/0303232].

  
  
\bibitem{peres} 
\art[hep-ph/0309312]{O.L.G. Peres, A.Yu. Smirnov}{Nucl. Phys.}{B680}{479}{2004}.


  
\bibitem{Feru}
 F.~Feruglio, A.~Strumia and F.~Vissani,
  Nucl.\ Phys.\ B {637} (2002) 345
  [addendum-ibid.\ B {659} (2003) 359]
  [hep-ph/0201291].
  
  

\bibitem{WMAP}
\hepart[astro-ph/0302207]{C.L. Bennett et al.}
and
\hepart[astro-ph/0302209]{D.N. Spergel et al.}.
The WMAP constraint  is similar to previous analyses, e.g.
\art{A. Lewis, S. Bridle}{\PR}{D66}{103511}{2002}.
Other (sometimes more conservative) analysis find similar (sometimes weaker) bounds:
\art[astro-ph/0303076]{S. Hannestad}{JCAP} {0305}{004}{2003},
\art[astro-ph/0306386]{S.W. Allen, R.W. Schmidt, S.L. Bridle}{Mon.\ Not.\ Roy.\ Astron.\ Soc.}{346}{593}{2003},
\art[astro-ph/0310723]{M.~Tegmark et al. [SDSS Collaboration]}{\PR}{D69}{103501}{2004}
\art[hep-ph/0312065]{V.Barger et al.}{\PL}{B595}{55}{2004},
\art[hep-ph/0402049]{P. Crotty, J. Lesgourgues, S. Pastor}{\PR}{D69}{123007}{2004},
\hepart[astro-ph/0407372]{U.~Seljak et al.}.

\bibitem{CosmoFuture}
For recent discussions see
  S.~Hannestad,  Phys.\ Rev.\ D {67} (2003) 085017 [astro-ph/0211106];
  J.~Lesgourgues, S.~Pastor and L.~Perotto,
   Phys.\ Rev.\ D {70} (2004) 045016 [hep-ph/0403296].


\bibitem{MAINZ}  MAINZ collaboration, hep-ex/0412056.

\bibitem{TROITSK}  TROITSK collaboration,
  Phys.\ Lett.\ B {460} (1999) 227.
  The latest results have been presented in a talk by V. Lobashev at the XI International Workshop on ``Neutrino Telescopes'' (Venezia, Feb. 22-25 2005).

\bibitem{Katrin} KATRIN web site: www-ik1.fzk.de/tritium.


  
  \bibitem{ig1} IGEX collaboration,
  Phys.\ Rev.\ C {59} (1999) 2108.
  
  
  \bibitem{cuo2} {\sc Cuoricino} collaboration, hep-ex/0501034.
  
  
\bibitem{HM}  Heidelberg-Moscow collaboration,
  Eur.\ Phys.\ J.\ A {12} (2001) 147
  [hep-ph/0103062].
  

\bibitem{staudt90}  A. Staudt {\em et al.}, Europh. Lett 13 (1990) 31.
\bibitem{staudt92} A. Staudt {\em et al.}, Phys. Rev. C46 (1992) 871. 
\bibitem{caurier96}  E.~Caurier, Phys. Rev. Lett. 77 (1996) 1954.
\bibitem{rodin03} V.A.~Rodin {\em et al.}, Phys. Rev. C68 (2003) 044302.

\bibitem{engel88} J.~Engel {\em et al.}, Phys. Rev. C37 (1988) 871. 

  
  \bibitem{ig2}  C.~E.~Aalseth {\it et al.},
  Mod.\ Phys.\ Lett.\ A {17} (2002) 1475
  [hep-ex/0202018].
  
  \bibitem{cuo1} {\sc Cuoricino} collaboration,
  Phys.\ Lett.\ B {584} (2004) 260.


\bibitem{MuraPenya}
H.~Murayama and C.~Pena-Garay,
  Phys.\ Rev.\ D {69} (2004) 031301 
emphasized that the lower bound on $|m_{ee}|$ in the case of inverted hierarchy
is `robust' and found a value in agreement with results in~\cite{Feru}.



  


\end{thebibliography}
\end{document}